\documentclass[manuscript,noacm]{acmart}
\AtBeginDocument{%
  }

\acmISBN{978-1-4503-XXXX-X/2018/06}

\usepackage[suppress]{color-edits} 
\usepackage{xspace}
\usepackage{color, colortbl}
\usepackage{xcolor}
\usepackage{enumitem}
\usepackage[singlelinecheck=false]{caption}
\usepackage{tabularx}  
\usepackage{multirow}
\usepackage{multicol}
\usepackage{threeparttable}
\usepackage{float}
\usepackage[bottom]{footmisc}
\usepackage{subcaption}
\usepackage{makecell}
\usepackage{adjustbox}
\usepackage{longtable}
\usepackage{booktabs}
\usepackage{array}
\usepackage{longtable}  
\usepackage{xcolor}
\usepackage{threeparttable}

\usepackage[most]{tcolorbox} 
\usepackage{fontawesome5}    
\usepackage{lmodern}

\usepackage{tcolorbox}
\usepackage{xcolor}

\tcbset{
  designguideline/.style={
    colback=gray!10,
    colframe=gray!60,
    boxrule=0pt,
    leftrule=4pt,
    arc=2pt,
    left=8pt,
    right=12pt,
    top=8pt,
    bottom=8pt,
    enhanced,
    before skip=8pt,
    after skip=8pt,
    width=\textwidth,
  }
}

\newtcolorbox{findingbox}[1][]{
    enhanced,
    colback=gray!5,
    colframe=gray!50,
    boxrule=0pt,
    leftrule=4pt,
    arc=2pt,
    left=10pt,
    right=10pt,
    top=8pt,
    bottom=8pt,
    before skip=8pt,
    after skip=8pt,
    breakable,
    #1
}

\newcommand{\GuidelineBox}[2]{%
  \begin{tcolorbox}[designguideline]
    \begin{tabular}{@{}p{1.2cm}@{}p{0.9\textwidth}@{}}
      \multirow{2}{*}{\centering {\fontsize{38}{38}\selectfont \faLightbulb[regular]}} 
        & \textbf{#1} \\
      & \rule{\linewidth}{0.4pt} \\
      & #2
    \end{tabular}
  \end{tcolorbox}
}

\newcolumntype{P}[1]{>{\centering\arraybackslash}m{#1}}
\newcolumntype{R}[1]{>{\raggedleft\let\newline\\\arraybackslash\hspace{0pt}}m{#1}}
\newcolumntype{L}[1]{>{\raggedright\let\newline\\\arraybackslash\hspace{0pt}}m{#1}}
\newcolumntype{C}{ >{\centering\arraybackslash} m{4cm} }

\definecolor{dawnblue}{rgb}{0.84, 0.92, 1.0}

\usepackage[capitalize,noabbrev]{cleveref}

\setlist[itemize]{align=parleft,left=0.5em..1.5em}

\newlist{req}{enumerate}{2}
\setlist[req,1]{label=RQ \arabic*:,ref= \textbf{\arabic*}, leftmargin=*}

\newlist{hyp}{enumerate}{2}
\setlist[hyp,1]{before=\itshape,font=\itshape, label=Hypothesis \arabic*:,ref= \arabic*, leftmargin=*}
\setlist[hyp,2]{before=\itshape,font=\itshape, label=Hypothesis \arabic*:,ref= \arabic*, leftmargin=*}

\begin{document}
\mathcode`(="4028
\mathcode`)="5029
\mathcode`+="202B
\mathcode`-="2200
\renewcommand{\sum}{\mathchar"1350}
\title{Believing vs. Achieving --- The Disconnect between Efficacy Beliefs and Collaborative Outcomes}

\author{Philipp Spitzer}
\email{philipp.spitzer@kit.edu}
\orcid{0000-0002-9378-0872}
\authornote{Both authors contributed equally to this research.}
\affiliation{%
  \institution{Karlsruhe Institute of Technology}
  \country{Germany}
}

\author{Joshua Holstein}
\email{joshua.holstein@kit.edu}
\authornotemark[1]
\affiliation{%
  \institution{Karlsruhe Institute of Technology}
  \streetaddress{Kaiserstr. 89}
  \city{Karlsruhe}
  \country{Germany}
  \postcode{76133}
}

\renewcommand{\shortauthors}{Spitzer et al.}

\begin{abstract}
\textit{As artificial intelligence (AI) becomes increasingly integrated into workflows, humans must decide when to rely on AI advice. These decisions depend on general efficacy beliefs, i.e., humans' confidence in their own abilities and their perceptions of AI competence. While prior work has examined factors influencing AI reliance, the role of efficacy beliefs in shaping collaboration remains underexplored. Through a controlled experiment (N=240) where participants made repeated delegation decisions, we investigate how efficacy beliefs translate into instance-wise efficacy judgments under varying contextual information. Our explorative findings reveal efficacy beliefs as persistent cognitive anchors, leading to systematic ``AI optimism''. Contextual information operates asymmetrically: while AI performance information selectively eliminates the AI optimism bias, data or AI information amplify how efficacy discrepancies influence delegation decisions. Although efficacy discrepancies influence delegation behavior, they show weaker effects on human-AI team performance. As these findings challenge transparency-focused approaches, we propose design guidelines for effective collaborative settings.}

\end{abstract}


\begin{teaserfigure}
\centering\includegraphics[width=.8\textwidth]{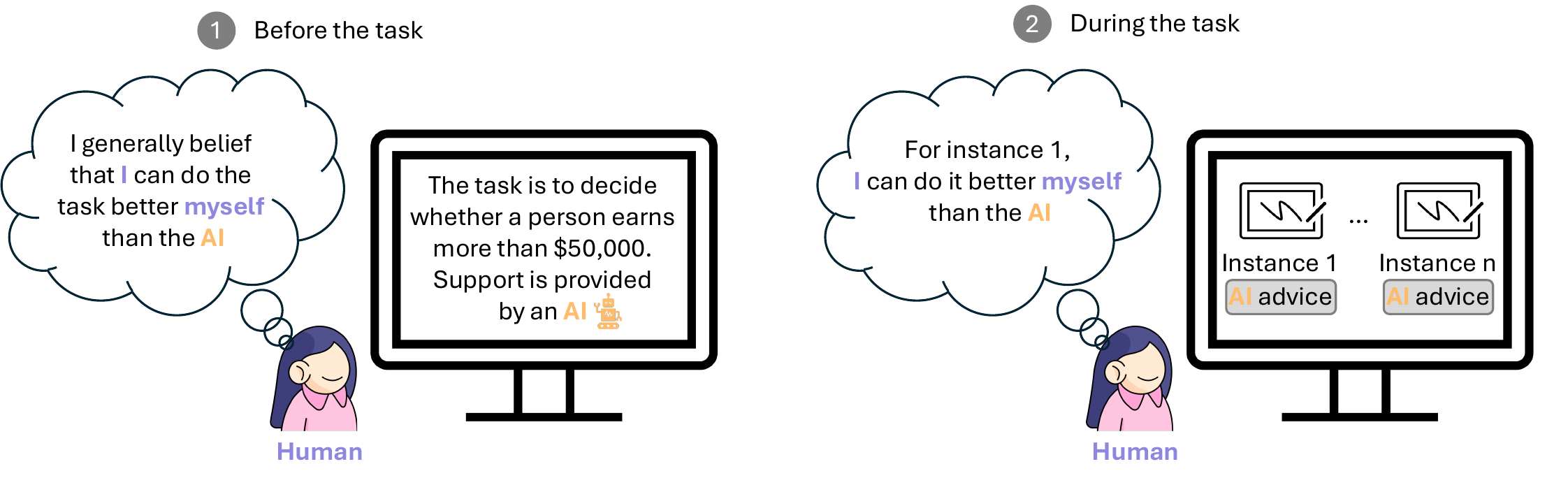}
\caption{The scheme of human-AI delegation and the impact of prior beliefs.}
\Description{Shows how a human has to decide whether to delegate a task to an AI or do it themselves. At the same time thinking about the efficacy of the AI and their own efficacy.}
\label{fig:teaser}
\end{teaserfigure}

\maketitle

\section{Introduction}
\label{sec-introduction}
In many domains, artificial intelligence (AI) is increasingly introduced to support humans with the promise of better decision-making outcomes. To live up to this, humans need to develop collaborative strategies to continuously decide whether to rely on their own assessment or follow the AI's advice \citep{li2023modeling, klingbeil2024trust, chen2023understanding}. For instance, content moderators must decide whether to follow AI-generated risk scores or to rely more on their own assessment \citep{gillespie2018custodians}, while financial analysts decide whether to trust their market domain expertise or rely on algorithmic trading recommendations \citep{lai2022human}. 

In these collaborative settings, two perceptions of competence fundamentally shape the reliance decisions. First, humans hold \textit{general efficacy beliefs}---perceptions of their own capabilities (self-efficacy) and the AI's competence (AI efficacy) that exist before engaging in specific tasks. Second, they make \textit{instance-wise efficacy judgments} when faced with particular decisions, evaluating their own ability and the AI's expected performance for that specific instance. These general beliefs and instance-wise judgments can influence one another: general efficacy beliefs can serve as cognitive anchors that bias subsequent instance-wise judgments \citep{bandura1977self, tversky1974judgment, goddard2004collective, nourani2021anchoring}. This anchoring effect reveals a critical tension in AI-assisted decision-making. While humans need accurate, context-sensitive, instance-wise judgments for effective delegation, their assessments may be systematically influenced by preexisting general beliefs that resist updating \citep{rastogi2022deciding}. Understanding how general efficacy beliefs translate into instance-wise judgments and how this translation affects collaboration outcomes is essential for designing AI that supports effective decision-making.

Recent research in human-computer interaction (HCI) has increasingly focused on scenarios in which humans and AI collaborate as teams, with humans often retaining decision-making authority over the final decision \citep{buccinca2024towards, vasconcelos2023explanations}. While numerous studies demonstrate the potential for AI integration to improve overall human-AI team performance \citep{bansal2019beyond, liu2021understanding}, substantial empirical evidence reveals that humans frequently exhibit overreliance on AI advice, i.e., following AI advice even when it is incorrect \citep{romeo2025exploring}.
Critically, overestimating AI competence can lead to automation bias, while low AI efficacy may result in missed opportunities to leverage AI strengths \citep{schemmer2023appropriate}. However, both forms of general efficacy beliefs do not necessarily need to align with actual ability; rather, they can function as cognitive filters that guide how humans interpret new task instances and make decisions \citep{spitzer2025human, pinski2023ai}. This disconnect between perceived and actual capabilities represents a core challenge in designing effective AI-assisted decision-making that enhances both human reliance strategies and team performance \citep{fernandes2024performance, ma2024you}.

Research has investigated the role of contextual information, showing its critical role in supporting human understanding and decision-making, including perceptions of one's own capabilities and those of the AI \citep{hogarth2007heuristic, mcrobert2013contextual, MAGNAGUAGNO2022102109}. This contextual information entails details about the decision environment that support humans in interpreting available information for their decisions \citep{Dey2001, brezillon1998using}. This, in turn, enables humans to calibrate their efficacy perceptions and decisions more effectively \citep{cai2019hello, cabrera2023improving}. Such contextual information includes, for example, information of the underlying data and its distribution \citep{Li2025}, local explanations revealing AI decision-making processes \citep{schoeffer2024explanations}, performance metrics of the AI across different scenarios \citep{Lu2021, He2023knowing}, or information on the AI's confidence \citep{zhang2020effect}. This information can calibrate human perceptions by providing reference points for judging AI strengths and limitations \citep{hemmer2022effect, pinski2023ai}, support more informed decision-making by revealing insights about the underlying domain \citep{Lai2020, schemmer2022influence}, and improve human understanding of when AI advice should be trusted or questioned~\citep{taudien2022calibrating, agarwal2023combining}.


Yet the relationship between contextual information and humans' efficacy perceptions is complex and not uniformly beneficial. For example, excessive contextual information that exceeds cognitive capacity can worsen decision-making \citep{shalu2025cognitive, peng2021does}, while poorly designed explanations can reinforce existing biases rather than correcting them \citep{morrison2024impact, spitzerimperfections}. 
This dual nature reveals a tension where contextual information is intended to help humans update their general efficacy beliefs appropriately, but instead increases cognitive biases or even creates new ones.
Understanding how contextual information influences the translation from general beliefs to instance-wise judgments and ultimately the collaboration outcomes is crucial for designing systems that effectively support human decision-making while maintaining appropriate levels of human agency and control \citep{lee2023leveraging}. Therefore, we pose the following research questions:

\begin{req}[leftmargin=1.06cm, labelindent=0pt, labelwidth=0em, label=\textbf{RQ\arabic*}:, ref=\arabic*]
    \item How do general efficacy beliefs anchor instance-wise efficacy judgments in AI-assisted decision-making?
    \label{rq1}
    \item How does contextual information influence the relationship between general efficacy beliefs and instance-wise judgments in AI-assisted decision-making?
    \label{rq2}
    \item How do discrepancies between general efficacy beliefs and instance-wise efficacy judgments influence collaboration behavior and, ultimately, AI-assisted decision-making?
    \label{rq3}
\end{req}

To address these questions, we conducted a controlled behavioral study for an income classification task. Participants (N=240) needed to decide whether to solve specific instances themselves or to delegate them to an AI under different contextual information: a control group without contextual information, a data group with information about the distribution of features in the dataset, an AI group with information about the AI's performance across those features, and a combined group with access to both types of information. We collected participants' general efficacy beliefs before task engagement and then had them continuously rate their instance-wise efficacy judgments during the study, allowing us to compare these instance-wise judgments against their general beliefs.

Our exploratory study reveals several critical insights into the role of efficacy perceptions in AI-assisted decision-making. First, we demonstrate that general efficacy beliefs serve as persistent cognitive anchors for subsequent instance-wise efficacy judgments, even when contextual information was provided. This anchoring effect proved particularly resistant for general self-efficacy beliefs. Second, we identified an ``AI optimism'' bias, in which participants systematically judged AI capabilities higher in specific instances than in their general efficacy beliefs.
Third, contextual information amplified how efficacy discrepancies influenced delegation behavior, though asymmetrically: data and AI information strengthened the relationship between self-efficacy discrepancies and control retention, while all information types amplified the relationship between AI efficacy discrepancies and delegation decisions.
Yet, these behavioral changes did not translate into performance improvements: efficacy discrepancies showed larger effects on delegation than on performance, suggesting that humans' intuitive instance-wise AI efficacy judgments systematically diverge from enhanced collaboration strategies. Thus, these findings advance the understanding of human factors in collaborative settings by showing the asymmetric nature of updates in efficacy (stable self-efficacy versus unstable AI efficacy) and revealing how contextual information can amplify rather than correct metacognitive biases. 

Our exploratory study makes three contributions to HCI: first, we develop a novel and human-centric research design based on established theory from behavioral science to link efficacy theory with human-AI collaboration. Second, we generate new insights into the anchoring role of efficacy perceptions and their interplay with delegation strategies. The findings challenge the value of purely transparency-based approaches in AI design. Third, we propose design guidelines for more effective collaborative settings.

\section{Related Work}
\label{sec-relatedwork}

In the following, we first clarify different concepts and establish the foundations of efficacy by drawing on literature from organizational psychology. We then present related work on the role of these in AI-assisted decision-making and how these are influenced through contextual information.

\subsection{Concepts}

While prior work has examined factors influencing AI reliance, such as general efficacy, confidence, and trust, the differences among these factors often remain opaque. These constructs represent distinct yet interrelated aspects of human cognition in collaborative settings. \Cref{tab:concepts} presents an overview that distinguishes these three constructs, with references to recent works that have taken up those concepts. In the following, we describe each concept in detail and clarify how our work is embedded within this landscape.

\begin{table*}[h]
\centering
\renewcommand{\arraystretch}{1.2}
\caption{Overview to distinguish different concepts: efficacy beliefs, confidence, and trust.}
\label{tab:concepts}
\begin{tabular}{p{0.32\textwidth}p{0.32\textwidth}p{0.32\textwidth}}
\toprule
 \textbf{Efficacy} & \textbf{Confidence} & \textbf{Trust} \\
\midrule
 
\cite{bandura1977self,bandura2006guide,hemmer2023human,Westphal2025,wu2024key,riefle2022may,ho2025defending} & 
\cite{li2025confidence,chong2023data,von2025knowing,ma2024you,ding2025new,hu2025being,koriat2012self,yeung2012metacognition} & 
\cite{de2020towards,kaplan2023trust,li2024developing,Yin2019,zhang2020effect,schemmer2023appropriate,lee1994trust,parasuraman2010complacency,dzindolet2003role,hoff2015trust,madhavan2007similarities,glikson2020human,jacovi2021formalizing,akash2020human,miller2021trust,mercado2016intelligent,chen2018situation, Kahr2024} \\
\midrule

Humans' assessments of their own capabilities (self-efficacy) and perceptions of AI competence (AI efficacy) to execute behaviors necessary to produce specific performance \cite{bandura1977self}. Efficacy can be measured as general beliefs formed before task engagement or as instance-wise judgments during actual task execution. General efficacy beliefs are relatively stable assessments formed through prior experience, social persuasion, and observational learning, while instance-wise efficacy judgments are more dynamic and context-dependent. & 
Subjective probability assessments about the likelihood of correctness for a specific decision or judgment. In AI-assisted decision-making, this encompasses both human self-confidence (one's certainty about their own judgment) and human confidence in AI (perceived likelihood that the AI's judgment is correct on a particular instance) \citep{li2025confidence, ma2024you}. Unlike efficacy, which assesses capability, confidence assesses expected correctness. Unlike trust, which operates at the system level, confidence is instance-specific. & 
Humans' perceptions of AI system reliability and their willingness to depend on the system \cite{lee1994trust,parasuraman2010complacency}. Appropriate trust---where perceived reliability matches actual system performance---enables effective collaboration by preventing both blind reliance (automation bias) and insufficient utilization (disuse). Trust develops through repeated interactions and performance feedback over time \cite{de2020towards}. \\
\bottomrule
\end{tabular}
\end{table*}

\textbf{Efficacy.} In organizational psychology, self-efficacy has long been recognized as a key determinant of motivation, effort, and performance. Grounded in Bandura's social cognitive theory~\cite{bandura1977self}, efficacy encompasses how humans evaluate both their own abilities (self-efficacy) and, in collaborative contexts, the capabilities of their partners. While high self-efficacy fosters persistence and task engagement, it can also reduce openness to seeking advice or relinquishing control; in contrast, low self-efficacy may encourage greater reliance on external support \citep{bandura2006guide}. In collaborative settings, general efficacy beliefs extend beyond personal capabilities to include assessments of partners' competence, influencing decisions about task allocation and when to seek or delegate assistance. A critical but often overlooked distinction exists between general efficacy beliefs---stable orientations formed through past experience, observation, and social influence---and instance-wise efficacy judgments made when confronting particular task instances. This distinction extends to AI-assisted decision-making, where humans must assess both their own capabilities (self-efficacy) and AI system capabilities (AI efficacy). Hemmer et al.\cite{hemmer2023human} found that allowing AI systems to delegate work to humans enhances human self-efficacy, subsequently improving both performance and satisfaction. This effect operated independently of explicit AI efficacy assessments, which remained implicit in delegation decisions. Westphal et al.\cite{Westphal2025} extended this understanding by demonstrating that self-efficacy moderates collaboration outcomes, with AI delegation particularly benefiting individuals with lower cognitive processing abilities through efficacy-boosting mechanisms. Pinski et al.~\cite{pinski2023ai} showed that providing humans with knowledge about AI capabilities during onboarding phases shapes both self-efficacy (through performance expectations) and AI efficacy (through task suitability appraisals), ultimately affecting delegation patterns. Despite these advances, existing research treats efficacy primarily as a pre-task construct, leaving unexplored how general beliefs dynamically translate into situation-specific evaluations during collaboration. 
In AI-assisted decision-making contexts, AI systems' opaque decision-making processes present particular challenges for efficacy assessment that distinguish them from human teammates \citep{ehsan2021explainable}. This opacity complicates humans' ability to accurately assess both their own and the AI's capabilities across contexts, potentially leading to miscalibrated collaboration strategies. While many studies measure one or more efficacy constructs, none explicitly examine how general efficacy beliefs formed before interaction translate into instance-wise efficacy judgments during actual task execution.

\textbf{Confidence.} Whereas efficacy assesses whether one possesses the capability to perform a task, confidence evaluates the perception of the likelihood that a specific decision will prove correct~\cite{koriat2012self}. Recent research has revealed confidence's dynamic nature in human-AI teams. \citet{li2025confidence} documented a critical confidence alignment phenomenon: displaying AI confidence systematically shifts human self-confidence, and this alignment persists even after the AI is removed. Through experimental studies, the authors documented bidirectional confidence alignment, where displaying AI confidence levels systematically shifts human self-confidence over repeated decisions. \citet{ma2024you} developed interventions to calibrate human confidence by presenting retrospective analyses of confidence-accuracy gaps, showing that metacognitive awareness improves appropriateness of reliance in subsequent decisions. A further study by \citet{von2025knowing} showed that explainable AI systems, while intended to support confidence calibration, can introduce new metacognitive distortions that undermine decision quality. These findings underscore confidence as an instance-level phenomenon responsive to immediate task features and AI behaviors, contrasting with the relatively stable nature of general efficacy beliefs.

\textbf{Trust.} Trust captures humans' expectations about system dependability and their resulting willingness to accept vulnerability when relying on AI~\cite{lee1994trust}. Appropriate trust calibration---where perceived reliability matches actual system performance---enables effective collaboration by preventing both excessive reliance (automation bias) and insufficient utilization (disuse)~\cite{parasuraman2010complacency}. \citet{hoff2015trust} synthesized 127 studies into a three-layer trust model distinguishing dispositional (personality-based), situational (context-dependent), and learned trust (experience-based). De Visser et al.~\cite{de2020towards} theorized trust as a longitudinally evolving construct shaped by cumulative interaction experiences, system transparency, and observed performance patterns across diverse contexts. Supporting this temporal perspective, \citet{Kahr2024} demonstrated that trust can be built up over repeated interactions with well-performing systems, making humans more tolerant of incidental AI errors that occur later versus earlier in the interaction. \citet{parasuraman2010complacency} identified attentional and cognitive mechanisms underlying trust miscalibration, showing how automation complacency emerges when humans shift monitoring resources away from reliable systems, making them vulnerable when those systems fail. \citet{dzindolet2003role} demonstrated that humans develop nuanced trust models incorporating system error types, performance boundaries, and contextual factors, with misunderstanding of these elements driving inappropriate reliance. Kaplan et al.'s~\cite{kaplan2023trust} meta-analysis synthesized these insights, revealing that trust formation depends on transparency, consistency, and domain characteristics, while distinguishing trust (attitude toward the system) from reliance (behavioral dependency). Unlike efficacy's focus on capability or confidence's emphasis on decision correctness, trust fundamentally concerns humans' willingness to depend on systems whose internal operations they may not fully comprehend.

These three constructs, i.e., efficacy, confidence, and trust, operate at different temporal and conceptual levels yet interact to shape collaboration outcomes. While Previous research has examined them largely in isolation: efficacy as stable pre-task orientations~\cite{hemmer2023human,Westphal2025}, confidence as dynamic instance-level assessments~\cite{li2025confidence,ma2024you}, and trust as accumulated reliance attitudes~\cite{de2020towards,parasuraman2010complacency}. Yet, there is no understanding of how general efficacy beliefs transform into context-sensitive instance-wise efficacy judgments when humans confront specific instances, and how this translation from general beliefs into instance-wise judgments influences collaborative behavior and performance. While our focus centers on efficacy beliefs, we situate this work within the broader landscape of related constructs measured in AI-assisted decision-making. \Cref{tab:related_work} provides an overview of recent research examining self-efficacy, AI efficacy, confidence, and trust. Notably, while many studies measure one or more of these constructs, none explicitly examine how general efficacy beliefs formed before interaction translate into instance-wise efficacy judgments during actual task execution.

\begin{table}[h!]
\centering
\footnotesize
\begin{threeparttable}
\caption{Summary of Related Work on Constructs Measured in AI-Assisted Decision-Making}
\Description{The table shows recent research divided by which constructs they measure: self-efficacy, AI efficacy, confidence, and trust.}
\label{tab:related_work}
\renewcommand{\arraystretch}{1.65}
\begin{tabular}{p{1cm}p{7.5cm}cccc}
\toprule
\textbf{Ref.} & \textbf{Summary} & \textbf{Self-Eff.} & \textbf{AI Eff.} & \textbf{Conf.} & \textbf{Trust} \\
\midrule

\citep{aman2025systematic} & 
Systematic review of human-AI collaboration literature. Emphasizes trust and reliability as central themes for AI efficacy but does not explicitly address self-efficacy. & 
× & o & o & $\checkmark$ \\

\citep{papantonis2023not} & 
Examines how AI uncertainty estimates and explanations affect human reliance and trust.  Measures instance-wise self-confidence but not general beliefs & 
o & × & $\checkmark$ & $\checkmark$ \\

\citep{ho2025defending} & 
Proposes a research agenda for explainable AI (XAI) to enhance cyber defense self-efficacy in distinguishing authentic from deepfake content. Discusses self-efficacy, but no empirical data have been collected yet. & 
o & × & × & o \\

\citep{Westphal2025} & 
Studies the impact of AI delegation on task performance and satisfaction. Identifies self-efficacy as a mediating mechanism, with visual processing ability moderating the indirect effect through self-efficacy. & 
$\checkmark$ & o & × & × \\

\citep{riefle2022may} & 
Investigates how general and instance-wise self-efficacy influences individuals' decisions to use conversational agents. & 
$\checkmark$ & × & × & × \\

\citep{taudien2024know} & 
Explores the relationship between metacognitive efficiency, confidence and human-AI collaboration performance. & 
× & × & $\checkmark$ & × \\

\citep{von2025knowing} & 
Examines how explainable AI affects human metacognition, specifically confidence calibration and delegation decisions.  & 
× & × & $\checkmark$ & × \\

\citep{ma2024you} & 
Explores how structured feedback on participants' past confidence ratings and corresponding accuracy helps calibrate human self-confidence. & 
× & × & $\checkmark$ & × \\

\citep{kim2024mental} & 
Examines AI adoption's impact on job stress and burnout. Self-efficacy in AI learning moderates the stress relationship. & 
$\checkmark$ & × & × & × \\

\citep{pinski2023ai} & 
Studies how AI knowledge affects delegation and task appraisal. Self-efficacy is implicit in human performance assessment, while AI efficacy is captured through task suitability appraisal. & 
o & o & × & × \\

\citep{li2025confidence} & 
Investigates how AI confidence influences humans' self-confidence and calibration in decision-making. Tracks self-confidence alignment with AI confidence as a dynamic process. & 
× & × & $\checkmark$ & × \\

\citep{ding2025new} & 
Proposes Bayesian model predicting trust behavior based on dual confidence measures. Explicitly measures both self-confidence and confidence in AI across varying difficulty levels. & 
× & × & $\checkmark$ & o \\

\citep{hu2025being} & 
Explores equal voting rights for AI agents and impact on team performance. Tracks changes in human self-confidence while varying AI efficacy through performance manipulation. & 
× & x & $\checkmark$ & o \\

\citep{chong2023data} & 
Dataset from chess puzzle study with AI assistance. Continuously measures both self-confidence and confidence in AI throughout task execution. & 
× & × & $\checkmark$ & × \\

\citep{Jeong2025} & 
Introduces a system designed to enhance users' self-efficacy in conversational AI interactions through visual progress feedback and subtask completion markers. & 
$\checkmark$ & × & x & × \\

\bottomrule
\end{tabular}
\begin{tablenotes}
\footnotesize
\item[] $\times$ = not measured, o = implicitly measured or discussed, $\checkmark$ = explicitly measured
\end{tablenotes}
\end{threeparttable}
\end{table}

\subsection{The Role of Contextual Information in Shaping Efficacy Beliefs}

Research in HCI highlights how contextual information shapes collaboration by helping humans form more accurate assessments of AI capabilities and potentially moderating the relationship between general beliefs and instance-wise judgments. This contextual information can be provided during initial onboarding phases or continuously throughout the collaboration, influencing how efficacy is formed and updated.

Onboarding and training phases represent critical opportunities to shape general efficacy beliefs before task execution begins. \citet{pinski2023ai} demonstrated that AI knowledge acquired during onboarding affects delegation efficiency and task appraisal alignment, with self-efficacy implicit in human performance assessment and AI efficacy captured through task suitability appraisal. Similarly, \citet{Lai2020} showed that onboarding with explanations and guidelines can enhance human performance in AI-assisted decision-making by helping humans develop more accurate initial efficacy assessments.

Information about the AI's general performance has been shown to significantly affect reliance decisions. \citet{Yin2019} demonstrated that communicating model accuracy affects both human reliance and perceived trustworthiness, while \citet{Kawakami2023} found that feedback containing information only available to humans improves detection of AI errors. Further,  \citet{Cai2019} showed that pathologists particularly value fine-grained performance measures that reveal AI strengths and weaknesses alongside contextual input details. \citet{cabrera2023improving} extended this by describing "model behaviors"—detailed descriptions of AI performance patterns across different contexts.

Contextual information about the underlying data and domain can also influence efficacy perceptions. \citet{holstein2023toward} found that providing information about otherwise unobservable data aspects substantially alters human interaction behavior. In medical contexts, \citet{agarwal2023combining} demonstrated that while AI predictions alone did not improve radiologists' diagnostic quality, access to clinical history data did enhance performance.

Some research explores the combined effects of AI and data information \citep{spitzer2025human}. \citet{Chiang2021} showed that allowing humans to inspect model performance disparities improves reliance calibration, though enabling assessment of both data and model performance did not necessarily improve outcomes. \citet{schemmer2022influence} conceptualized this in terms of information asymmetry, arguing that optimal performance requires humans and AI to access different but compatible contextual information sources.

To summarize, while existing research has examined efficacy beliefs (either as static pre-task assessments or dynamically during task execution), confidence (as instance-level assessments), and trust (as accumulated attitudes), none explicitly examine how general efficacy beliefs formed before interaction translate into instance-wise efficacy judgments during the actual task. Our work investigates this relationship between general beliefs and instance-wise judgments, examining how contextual information moderates this translation and how discrepancies between these two levels of efficacy beliefs influence delegation behavior and ultimately shape human-AI team performance.
\section{Methodology}

\subsection{Experimental Design}
We conducted a behavioral study to examine how general efficacy beliefs align with instance-specific judgments in human-AI delegation, and how contextual information moderates this relationship. In the following, we describe the experimental design, including the task and the AI, as well as our treatments and the overall procedure. The study was approved by the Institutional Review Board (IRB).

\textbf{Task and Data.} We selected an income classification task using the American Community Survey dataset from \citet{ding2021retiring}, which is based on the ACS Public Use Microdata Sample (PUMS). The dataset consists of 1,599,229 instances with ten features and one target variable representing whether US citizens earn above or below \$50,000 annually. Similar to previous studies \citep{bordt2022post, ezzeldin2023fairfed}, we chose this task because of its accessibility for laypeople with limited domain expertise. 

\textbf{Feature Selection and Pre-Testing.} To avoid cognitive overload \citep{paleja2021utility, dietzmann2022artificial}, we aimed to reduce our feature set while maintaining sufficient information for informed decisions. Therefore, we trained a decision tree on the data and selected the four features with the largest feature importance: \textit{age}, \textit{education}, \textit{occupation}, and \textit{hours worked per week in the past 12 months}. We validated this feature reduction through pre-testing, where participants achieved performance levels comparable to previous studies that display all features to participants \citep{chen2023understanding}.

\textbf{AI System and Instance Selection.} We implemented a decision tree classifier as the AI. We randomly partitioned the dataset into a training set (70\%) and a validation set (30\%) on which the AI achieved an accuracy of 77\%. From the validation set, we randomly sampled 60 instances for the study.

\textbf{Collaboration Type.} To implement a collaborative setting between humans and AI, two distinct paradigms have emerged: decision support, where the AI provides advice that humans can accept or reject, and delegation, where task instances are transferred between humans and AI systems \citep{hemmer2024complementarity}. Two complementary directions of delegation have been studied: (i) the AI delegating task instances to humans \citep{fugener2022cognitive, hemmer2023human, raghu2019algorithmic}, and (ii) humans delegating task instances to the AI \citep{lubars2019ask, pinski2023ai}. While both lines of research reveal important insights, our focus lies on the latter: how humans decide to entrust AI with specific instances and the implications of this delegation behavior. By reflecting the capabilities of both themselves and the AI and deciding who is more suitable to solve the instance, it provides a structured setting for studying the effects of general efficacy beliefs and instance-wise efficacy judgments. Consequently, we implement AI-assisted decision-making as a delegation setting, where humans can delegate a task instance to an AI. The findings can also be transferred to collaboration settings, in which the AI provides decision support.

\textbf{Treatments.} Our between-subjects experiment ($N = 240$) uses a 2×2 factorial design with two varying factors: i) the availability of data distribution information and ii) the availability of AI performance information. 
This design directly addresses our research questions by systematically manipulating contextual information types that prior work has identified as influential for AI-assisted decision-making~\cite{holstein2023toward,schemmer2022influence}. Data distribution information provides humans with domain knowledge about the underlying task structure, potentially affecting instance-wise self-efficacy judgments by revealing task characteristics and patterns~\cite{holstein2023toward}. AI performance information makes system capabilities transparent by revealing the AI's strengths and limitations across different contexts. This transparency can shape instance-wise AI efficacy judgments~\cite{Cai2019,cabrera2023improving}, which has been shown to calibrate human reliance on AI advice \citep{Lu2021, He2023knowing}. The control condition (no contextual information) enables us to examine the baseline relationship between general efficacy beliefs and instance-wise judgments. By independently varying these two information types, we can examine whether contextual information moderates the relationship between general efficacy beliefs and instance-wise judgments, with the factorial design allowing us to distinguish effects of domain knowledge (data information) from effects of AI knowledge (AI information), as well as potential interaction effects when both are available. Finally, by measuring delegation decisions and performance outcomes across all conditions, we can investigate how efficacy discrepancies influence collaboration behavior and team performance, and whether these relationships differ depending on available contextual information.
This results in four conditions with access to different types of contextual information: control (no additional information), only contextual data information (visualizations of feature distributions with respect to income), only contextual AI information (visualizations of feature distributions with respect to the AI's accuracy), and combined (contextual data and AI information available). Illustrations of this information can be found in \Cref{fig: contextualinformation} in the Appendix.

\begin{figure}[h!]
    \centering
    \includegraphics[width=1\linewidth]{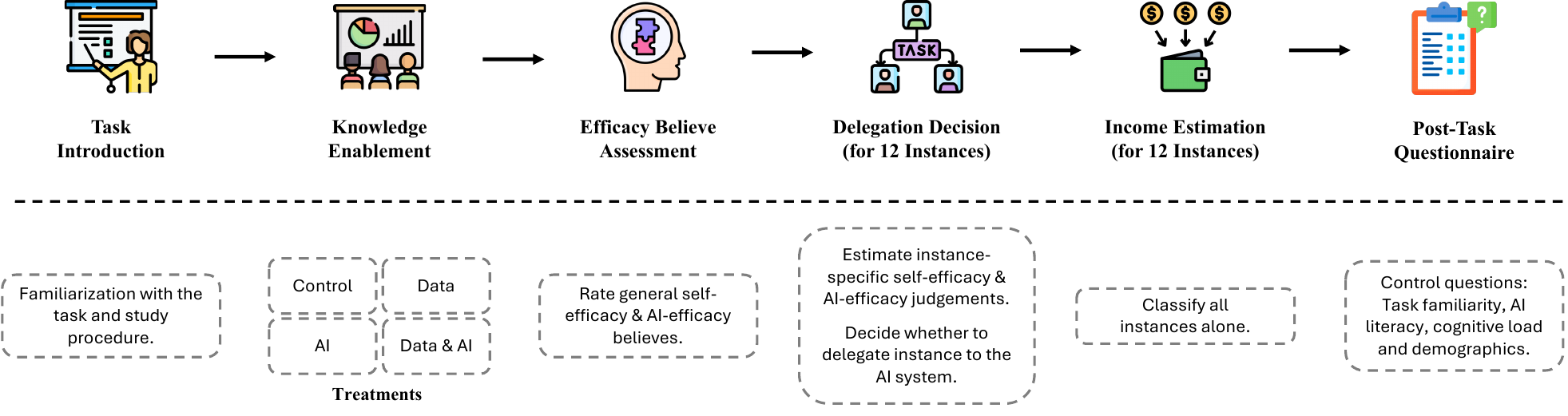}
    \caption{Experimental design examining how contextual information influence general efficacy beliefs and instance-wise efficacy judgments, and how these relationships translate into delegation decisions in AI-assisted decision-making.}
    \Description{The figure shows the different phases of the study conducted in this work.}
    \label{fig:procedure}
\end{figure}

\textbf{Procedure.} The procedure of our study was set up in six sequential phases (see \Cref{fig:procedure}). In the beginning, participants were randomly assigned to one of the four conditions: control, contextual data information, contextual AI information, or both:

\begin{enumerate}
    \item \textbf{Task Introduction.} In the beginning, participants provided consent to our study and received an introduction to the task. This was followed by an explanation of the study procedure, which established a baseline understanding before encountering specific instances that might influence general efficacy beliefs.
    
    \item \textbf{Knowledge Enablement.} Participants were then made familiar with their assigned contextual information, i.e., data distributions, AI performance across feature distributions, or both types of information. Control participants received no additional information. To ensure engagement with the provided information, participants were required to view all available contextual information before proceeding for at least 60 seconds. Following this exposure, participants completed a comprehension check to verify their understanding of the available information.
    
    \item \textbf{Efficacy Belief Assessment.} Before the task, participants rated their general self-efficacy beliefs as well as their general AI efficacy beliefs for correctly classifying people's income using the established scale of \citet{spreitzer1995psychological}. These ratings serve as our measures of \textit{general efficacy beliefs}.
   
    \item \textbf{Delegation Decision.} In the main phase of the study, participants received twelve randomly selected instances from a pool of 60 instances. This number was chosen to manage study length and cognitive load while ensuring sufficient within-subject variation, aligning with previous human-AI decision-making studies \citep{buccinca2025contrastive, schoeffer2024explanations}. For each instance, participants rated their \textit{instance-wise self-efficacy judgments and AI efficacy judgments} and made a binary choice of whether to delegate the task instance to the AI or classify it themselves. Inspired by previous research \citep{oppenheimer2009instructional, Lai2020}, we imposed a 10-second minimum viewing time before being able to proceed to the next instance to ensure sufficient engagement with the task instance and the contextual information. Participants had access to their assigned contextual information at any time during this phase by clicking a button. Importantly, participants received no feedback on the correctness of their delegation decisions or the AI's performance on specific instances during this phase, ensuring that instance-wise efficacy judgments remained unaffected by outcome knowledge.
    
    \item \textbf{Income Estimation.} To obtain human-alone performance as a baseline, participants then classified all twelve instances in a randomized order themselves, regardless of their previous delegation decisions. 
    
    \item \textbf{Post-Task Questionnaire.} Finally, participants completed a post-task questionnaire where we collected data on demographic information and allowed participants to provide open feedback.

\end{enumerate}

During the \textit{Delegation Decision} and \textit{Income Estimation} phases, participants did not see the AI's predictions for specific instances. This design ensured that instance-wise efficacy judgments reflected participants' perceptions based solely on instance features and contextual information, rather than being influenced by performance feedback. The contextual information that was shown to participants can be seen in the Appendix in \Cref{fig: contextualinformation}. The design of the contextual information aligns with \citet{cabrera2023improving}.

\textbf{Measurement.}
Our analysis centers on efficacy alignment by comparing general self-efficacy and AI efficacy beliefs with instance-wise efficacy judgments. To obtain the general self-efficacy and AI efficacy beliefs, participants rated three items on a seven-point Likert scale from ``Strongly disagree''to ``Strongly Agree'' based on the established items of \citet{spreitzer1995psychological}. 
For each instance, participants rated their instance-wise self-efficacy and AI efficacy judgments on a slider ranging from ``0\% (not at all)'' to ``100\% (well suited)'' by answering the following questions, which were adapted from \citet{pinski2023ai}: ``How well are you suited to solve this task?'' and ``How well is the AI suited to solve this task?'' respectively.

Further behavioral outcomes include delegation rate, measured as the proportion of instances delegated to the AI \citep{fuegener2021exploring, hemmer2023human}, and delegation performance (the human-AI team performance based on delegation decisions). A correct delegation decision occurs when participants delegate to the AI and the AI's prediction matches the ground truth, or when participants choose to solve it themselves and their own classification is correct. We assessed AI knowledge using the scale developed by \citet{ehsan2024xai}. Task familiarity was measured with a single item (``How familiar are you with the task domain of classifying people's income?'') on a six-point Likert scale ranging from ``Not familiar'' to ``Very familiar''. An overview of all items is provided in \Cref{tab:all_measures} in the Appendix.

\textbf{Participants.}
We recruited 60 participants per treatment, resulting in  240 participants in total (128 female, 110 male, 2 unspecified;$ \mu_{age} = 39.4, \sigma_{age} = 12.76$, median completion time: 20.42 minutes) through Prolific.co from English-fluent US residents, yielding 2,880 observations for analysis. Throughout the study, we implemented multiple quality control measures as recommended by prior research \citep{abbey2017attention}. Participants in treatment conditions completed two comprehension check questions following their exposure to contextual information; only participants who answered at least one of the two questions correctly were included in the sample: 176 participants (97.8\%) answered both questions correctly, while 4 participants (2.2\%) answered only one correctly. Participants also had to successfully pass at least one out of two attention checks. Participants who successfully finished the study received \pounds2.25 base compensation in addition to performance-based bonuses: 5 pennies per correct delegation decision (when a delegated task was correctly solved by the AI or alternatively, if the human classified the instance correctly) and an additional 5 pennies per correctly classified instance in the income estimation phase. Participants reported, on average, low task familiarity (M = 2.52, SD = 1.27 on a 6-point scale) and low AI knowledge (M = 2.56, SD = .9 on a 5-point scale), with no significant differences across conditions (F(3,236) = .582, p = .627 and F(3,236) = 1.01, p = .39, respectively).

\subsection{Framework for Efficacy-Based Delegation}
\label{subsec-framework}
We formalize the relationship between efficacy and delegation behavior in the following.
We model instance-wise efficacy judgments as functions of general beliefs and instance characteristics. Each task instance $i$ is characterized by a feature vector $x_i \in \mathcal{X}$, where $\mathcal{X}$ represents the feature space containing relevant task characteristics. For human $j$, let $S_j^{b} \in \left[0,1\right]$ denote their self-efficacy belief and $A_j^{b} \in \left[0,1\right]$ denote their general AI efficacy belief, measured before task engagement. 
For a specific task instance $i$ with features $x_i$, human $j$ forms instance-wise efficacy judgments that we model as:
\begin{align}
S_{i,j}(x_i) &= \alpha_s * S_j^{b} + (1 - \alpha_s) * f_s(x_i) \\
A_{i,j}(x_i) &= \alpha_a * A_j^{b} + (1 - \alpha_a) * f_a(x_i)
\end{align}
where $f_s(x_i)$ and $f_a(x_i)$ represent feature-dependent functions that capture how the specific characteristics of task instance $i$ influence self-efficacy and AI efficacy judgments, respectively. The parameters $\alpha_s$ and $\alpha_a$ capture the anchoring effect of general beliefs on instance-wise judgments.


For each instance $i$, human $j$ has discrepancies between their instance-wise judgments and general beliefs:
\begin{align}
\Delta S_{i,j} &= S_{i,j} - S_j^{b} \quad \text{(self-efficacy discrepancy)} \\
\Delta A_{i,j} &= A_{i,j} - A_j^{b} \quad \text{(AI efficacy discrepancy)}
\end{align}
These discrepancies capture how humans adapt their general beliefs for specific task instances. The probability that human $j$ delegates instance $i$ follows:
\begin{equation}
P(\text{delegate}_{i,j}) = \sigma(\beta_0 + \beta_s \Delta S_{i,j} + \beta_a \Delta A_{i,j})
\end{equation}
where $\sigma(\cdot)$ is the logistic function. The coefficients have the following interpretations: $\beta_0$ represents the baseline tendency to delegate,  $\beta_s$ captures how self-efficacy discrepancies influence delegation probability, and  $\beta_a$ captures how AI-efficacy discrepancies influence delegation probability. Positive values indicate that larger discrepancies increase the probability of delegation, while negative values indicate the opposite.

\section{Results}
\label{sec-results}

The results section is organized around our three research questions: first, we analyze discrepancies in participants' general efficacy beliefs and instance-wise efficacy judgments across conditions. 
Second, we investigate how contextual information moderates this efficacy discrepancy. 
Third, we analyze how those efficacy discrepancies affect delegation behavior and human-AI team performance. 
Throughout all analyses, we report coefficients with standard errors and apply significance thresholds of $p < .05$, $p < .01$, and $p < .001$. Additionally, we report effects with $p < .10$ as non-significant trends.

\subsection{RQ1 - Relationship of General Efficacy Beliefs and Instance-Wise Efficacy Judgments}


To examine how general efficacy beliefs anchor instance-wise efficacy judgments (RQ\ref{rq1}), we calculate average efficacy discrepancies per participant as the difference between instance-wise judgments and general beliefs. Positive values indicate participants rated their instance-wise efficacy higher than their general beliefs, while negative values indicate the opposite. Specifically, we test the existence of these discrepancies by using one-sample t-tests that test for differences from zero. To account for multiple comparisons, we applied the Benjamini-Hochberg false discovery rate correction separately for self-efficacy (four tests) and AI efficacy (four tests). We report corrected q-values alongside uncorrected p-values. Assumption checks confirmed homogeneity of variances across conditions (Levene's test: self-efficacy $W=1.18, p=.319$; AI efficacy $W=.16, p=.924$). While Shapiro-Wilk tests revealed mild deviations from normality for self-efficacy discrepancies in the combined condition ($p=.0286$) and for AI efficacy discrepancies in the AI condition ($p=.0138$), as well as the combined conditions ($p=.027$), the large sample size (N=60 per condition) provides robustness for parametric tests (for full details see \Cref{tab:normality-pvals}). Non-parametric Wilcoxon signed-rank tests yielded consistent conclusions, confirming the reliability of our findings.

\begin{figure}[h!]

\begin{subfigure}{.45\textwidth}
  \centering
  \includegraphics[width=\textwidth]{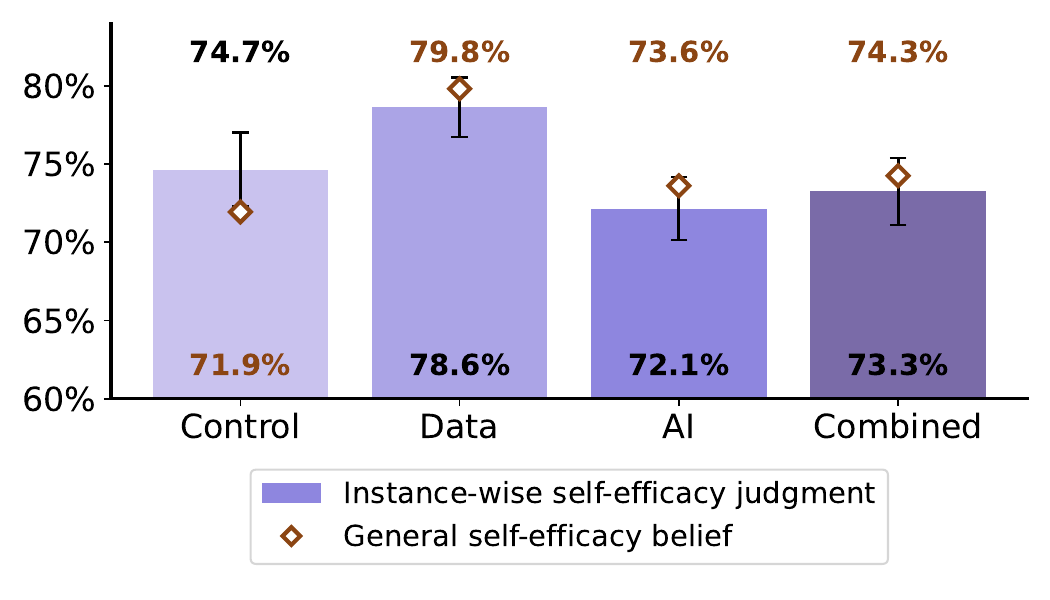}
    \captionsetup{justification=centering}
    \label{fig: self-efficacy-barplots}
  \caption{Self-efficacy.}
\end{subfigure}
\begin{subfigure}{.45\textwidth}
  \centering
  \includegraphics[width=\textwidth]{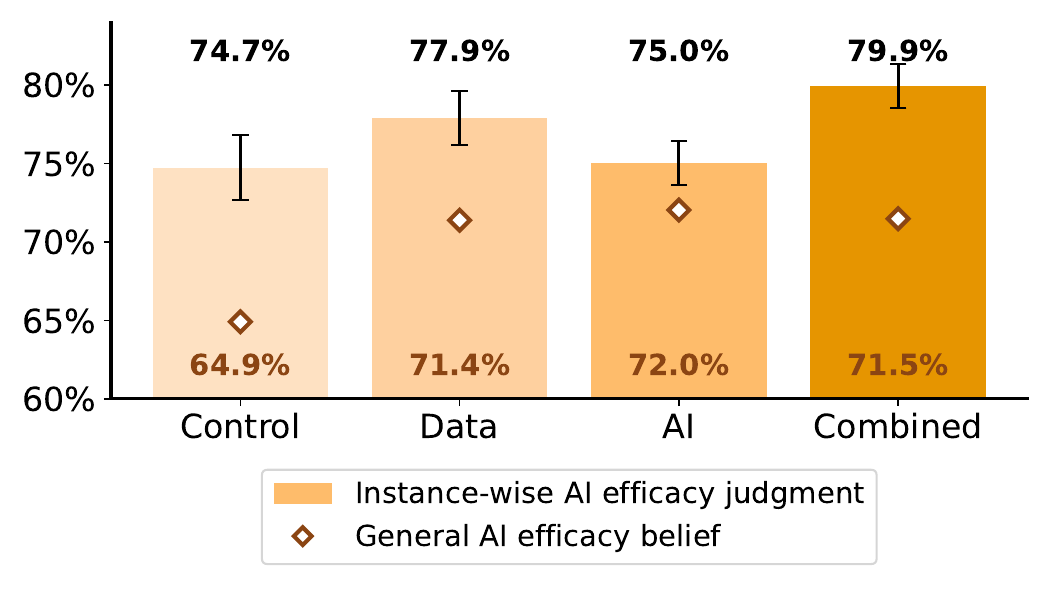}
    \captionsetup{justification=centering}
    \label{subfig: correct}
  \caption{AI efficacy.}
  \centering
\end{subfigure}

\captionsetup{justification=raggedright,singlelinecheck=false}
\caption{General efficacy beliefs and instance-wise efficacy judgments for self-efficacy and AI efficacy. AI efficacy discrepancies are larger than for self-efficacy.}
\Description{The plots show the general and instance-wise self-efficacy and AI efficacy as barplots and lines in two subfigures.}
\label{fig: efficacy-barplots}
\end{figure}

\textbf{Self-Efficacy: Well-Calibrated Instance-Wise Judgments.} \Cref{fig: efficacy-barplots}a shows that general self-efficacy beliefs ranged between 72-80\% across conditions, closely aligned with instance-wise judgments, which are slightly lower (72-79\%). Testing whether discrepancies between general beliefs and instance-wise judgments differed from zero revealed no significant deviations in any condition: Control ($\Delta =.027, p=.220, q=.654, d=.16$), Data ($\Delta =-.012, p=.475, q=.654, d=-.09$), AI ($\Delta =-.015, p=.491, q=.654, d=-.09$), and Combined ($\Delta =-.010, p=.682, q=.682, d=-.05$). All effect sizes were negligible ($|d| < .2$), indicating that participants' instance-wise self-efficacy judgments aligned with their general beliefs. Self-efficacy beliefs functioned as accurate cognitive anchors rather than sources of systematic bias (for full results see \Cref{tab:self-efficacy-discrepancy}).

\textbf{AI Efficacy: Systematic Instance-Wise Optimism.} \Cref{fig: efficacy-barplots}b shows that general AI efficacy beliefs ranged between 65-72\%, with instance-wise judgments consistently exceeding these beliefs across all conditions, ranging from 68\% to 75\%. This systematic upward shift created positive discrepancies in all conditions, though the magnitude varied by contextual information. The Control condition showed a significant medium-sized AI optimism bias ($\Delta =.098, p<.001, q<.001, d=.59$), indicating that participants without contextual information systematically overestimated AI capabilities by approximately 10 percentage points when evaluating specific instances. The Data condition showed a lower bias ($\Delta = .065, p=.001, q=.001, d=.45$), showing a small yet significant effect. Similarly, the Combined condition retained significant optimism ($\Delta =.085,  p<.001, q<.001, d=.49$). Critically, the AI condition was the only one where discrepancies became non-significant ($\Delta =.030, p=.151, q=.151, d=.19$), with a negligible effect size. Solely providing information on AI performance uniquely eliminated the systematic optimism bias that persisted across all other conditions (for full results see \Cref{tab:ai-efficacy-discrepancy}).

\begin{findingbox}
    \textbf{Finding 1:} General efficacy beliefs anchor instance-wise judgments asymmetrically. Self-efficacy shows alignment across conditions. AI efficacy shows systematic optimism, which Data and Combined conditions reduce but do not eliminate. Only AI information eliminates this optimism bias.
\end{findingbox}

\begin{figure}[h!]
  \centering
  \includegraphics[width=.65\textwidth]{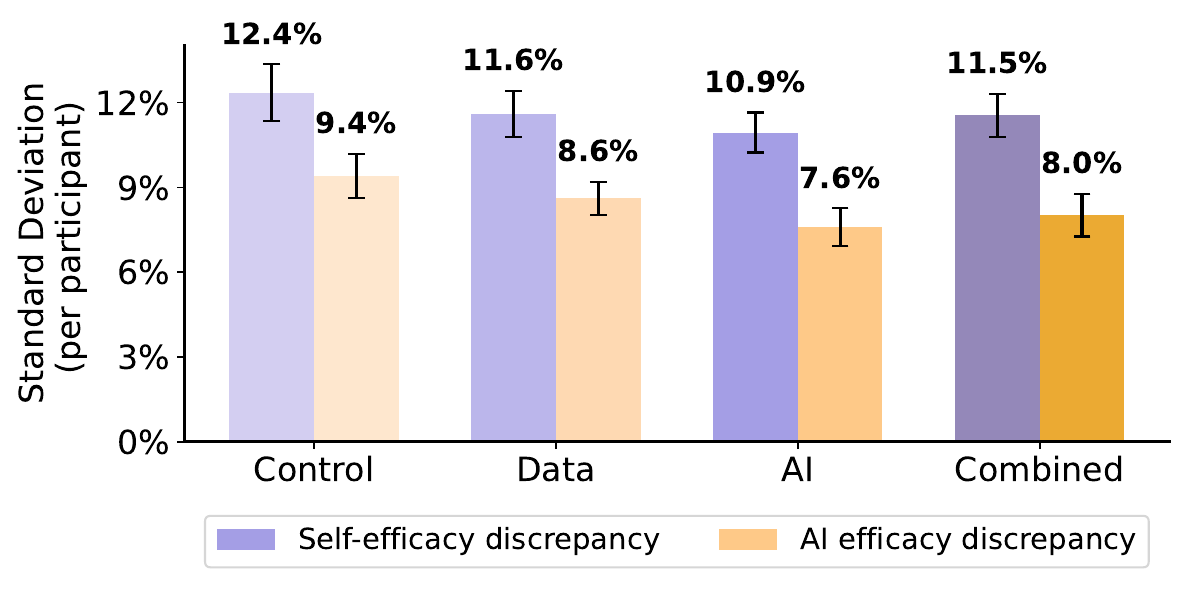}
    \captionsetup{justification=centering}
    \centering
\captionsetup{justification=raggedright,singlelinecheck=false}
\caption{Instance-wise deviations for self-efficacy and AI efficacy.}
\Description{Box-plots are shown for instance-wise self-efficacy and AI efficacy.}
\label{fig: instance-based deviations}
\end{figure}

To further examine the consistency of efficacy discrepancies across instances, we calculated the standard deviation of each participant's discrepancies across the twelve instances. We tested whether self-efficacy and AI efficacy discrepancies showed different levels of within-person variability using a paired Wilcoxon signed-rank test, given violations of normality in the difference scores ($W=.913, p < .001$).
\Cref{fig: instance-based deviations} shows that self-efficacy discrepancies varied more than AI efficacy discrepancies across all conditions. The Wilcoxon signed-rank test confirmed this pattern: self-efficacy discrepancies showed significantly greater within-person variability ($\mu =.116$) compared to AI efficacy discrepancies ($\mu=.084, W=5036, p<.001, d=-.48$). This pattern held consistently across all conditions (all p<.01). While general self-efficacy beliefs aligned with average instance-wise judgments (as shown in Finding 1), participants varied in their instance-wise judgments. In contrast, AI efficacy discrepancies remained more consistent, reflecting a stable, systematic instance-wise AI optimism bias.

\begin{findingbox}
    \textbf{Finding 2:}
    Self-efficacy discrepancies show greater instance-to-instance variability than AI efficacy discrepancies, indicating more flexible adjustment for self-assessment.
\end{findingbox}

\subsection{RQ2 - Contextual Information on General Efficacy Beliefs and Instance-Wise Judgments}

To address \textbf{RQ\ref{rq2}}, how contextual information influences the relationship of general efficacy beliefs on instance-wise efficacy judgments, we conduct two OLS regression analyses (see \Cref{tab: regression-moderation}). Here, we model instance-wise efficacy judgments as the dependent variable, general efficacy beliefs as the independent variable, and contextual information as the moderating variable.
The model on self-efficacy judgments explains 34.9\% of the variance ($R^2 = .349$, $F = 17.77$, $p < .001$). General self-efficacy belief emerges as a significant positive predictor ($coef = .552$, $p < .001$), indicating that participants with higher general self-efficacy beliefs also rated themselves more favorably on individual instances. However, none of the contextual information conditions significantly moderate this relationship, suggesting that general self-efficacy beliefs remain relatively stable predictors of instance-wise judgments regardless of available information.

For instance-wise AI efficacy judgments, the model accounts for 21.9\% of the variance ($R^2 = .219$, $F = 9.295$, $p < .001$). General AI efficacy belief emerges as a significant positive predictor ($coef = .470$, $p < .001$), confirming that general beliefs about AI competence are carried forward to instance-wise judgments. The interaction with combined contextual information shows a significant negative effect ($coef = -.367$, $p = .007$) (see \Cref{tab: regression-moderation}), indicating that when having access to both data and AI information, participants' general AI efficacy beliefs have a smaller impact on their instance-wise AI efficacy judgments (see visualized interaction effect in the Appendix in \Cref{fig: aiefficacy-interaction}). This interaction effect suggests that comprehensive contextual information reduces reliance on general beliefs, enabling more instance-specific AI efficacy judgments.

\begin{findingbox}
    \textbf{Finding 3:} General efficacy beliefs predict instance-wise judgments during AI-assisted decision-making. 
\end{findingbox}

In human-AI delegation contexts, this moderating effect indicates that providing rich contextual information may help humans to make more situationally specific decisions about when to rely on AI assistance rather than defaulting to their pre-existing assumptions about AI capabilities.

\begin{findingbox}
    \textbf{Finding 4:} Self-efficacy judgments remain stable across information conditions, whereas AI efficacy judgments become less dependent on general beliefs when both data and AI information are available. 
\end{findingbox}

\begin{table}[h!]
\caption{Regression Results for Instance-Wise Efficacy Judgments by Condition}
\Description{The table shows the results of the regression for Efficacy Judgments by Condition.}
\label{tab: regression-moderation}
\begin{center}
\begin{tabular}{lcccc}
\toprule
  & \multicolumn{4}{c}{Instance-Wise Efficacy Judgments} \\
\cmidrule(lr){2-5}
                                                         & \multicolumn{2}{c}{Self-Efficacy} & \multicolumn{2}{c}{AI Efficacy} \\
\cmidrule(lr){2-3} \cmidrule(lr){4-5}
                                                         & coef & std err & coef & std err \\
\midrule
Intercept                                        & \textbf{.350***} & .060 & \textbf{.442***} & .061 \\
Condition: Data                      & -.067 & .115 & .023 & .090 \\
Condition: AI                        & .028 & .095 & .141 & .094 \\
Condition: Combined                     & .092 & .093 & \textbf{.284**} & .095 \\
General efficacy belief                             & \textbf{.552***} & .080 & \textbf{.470***} & .091 \\
General efficacy belief $\times$ Data & .080 & .144 & -.029 & .129 \\
General efficacy belief $\times$ AI  & -.085 & .125 & \textbf{-.238$^\dagger$} & .133 \\
General efficacy belief $\times$ Combined & -.160 & .123 & \textbf{-.367**} & .136 \\
\midrule
R-squared & \multicolumn{2}{c}{.349} & \multicolumn{2}{c}{.219} \\
F-statistic & \multicolumn{2}{c}{17.77***} & \multicolumn{2}{c}{9.295***} \\
\bottomrule
\end{tabular}
    \begin{tablenotes}
        \item[1] Note: \textit{$^\dagger$p < .10; *p < .05; **p < .01; ***p < .001}
    \end{tablenotes}
\end{center}
\end{table}

\subsection{RQ3 - General Efficacy Beliefs and Instance-Wise Judgments on Delegation and Performance}

To address \textbf{RQ\ref{rq3}}, we examine how discrepancies between instance-wise judgments and general efficacy beliefs influence delegation behavior and human-AI team performance. We analyze this relationship at the instance level using two binary mixed-effects regression models (see \Cref{tab: regression-mixedeffects}). We model the delegation behavior and human-AI team performance as dependent variables, the discrepancies between general efficacy beliefs and instance-wise judgments as independent variables, contextual information as a moderating variable, and set random intercepts for participant and instance IDs.

The analysis of delegation behavior reveals substantial main effects for both efficacy discrepancies. Self-efficacy discrepancies show a large negative effect on delegation behavior ($coef = -3.066$, $p < .001$), indicating that when participants rate their own capabilities lower for a specific instance compared to their general efficacy belief, they are substantially more likely to delegate the instance to the AI. AI efficacy discrepancies further show a moderate positive effect ($coef = 1.119$, $p < .001$), meaning that when participants rate the AI more favorably on specific instances compared to their general AI beliefs, they become more willing to delegate.

The interaction effects reveal how contextual information alters the efficacy-delegation relationships. For self-efficacy discrepancies, the Data condition shows a significant interaction effect ($coef = -1.549$, $p < .001$). The AI condition shows a moderate interaction ($coef = -1.112$, $p = .009$), while the Combined condition shows no significant effects ($coef = -.645$, $p = .106$). This means that access to either data or AI information increases the effect of self-efficacy and AI efficacy discrepancies: when contextual information is available, humans are more likely to delegate when they rate their instance-wise self-efficacy lower than their general efficacy belief. When they rate themselves higher than originally, the effect of retaining control increases. For AI efficacy discrepancies, all conditions show positive interaction effects of varying magnitudes, and the AI condition has the largest interaction effect: Data ($coef = .871$, $p = .016$), AI ($coef = 2.202$, $p < .001$), and Combined ($coef = 1.173$, $p = .001$). This means that having access to contextual information increases the effect of efficacy discrepancy on the likelihood to delegate: when having access to contextual information, participants delegate more often when they show higher AI efficacy ratings for specific instances compared to their general beliefs.

\begin{findingbox}
    \textbf{Finding 5:} Contextual information creates an asymmetric amplification effect: data and AI information amplify control retention when instance-wise self-efficacy judgments exceed general beliefs, while all information types amplify delegation when instance-wise AI efficacy judgments exceed general beliefs.
\end{findingbox}

\begin{table}[h!]
\caption{Mixed-Effects Logistic Regression Results for Delegation Behavior and Performance. Coefficients are given in log-odds.}
\Description{The table shows the results of the regression for Delegation Behavior and Performance. Coefficients are given in log-odds.}
\label{tab: regression-mixedeffects}
\begin{center}
\begin{tabular}{lcccc}
\toprule
& \multicolumn{2}{c}{Delegation} & \multicolumn{2}{c}{Performance} \\
\cmidrule(lr){2-3} \cmidrule(lr){4-5}
& coef & std err & coef & std err \\
\midrule
Intercept                                        & \textbf{-1.810***} & .295 & \textbf{1.220***} & .266 \\
Self-efficacy discrepancy                     & \textbf{-3.066***} & .264 & \textbf{-.359***} & .106 \\
AI efficacy discrepancy                          & \textbf{1.119***} & .196 & \textbf{.162$^\dagger$} & .090 \\
Condition: Data                                  & \textbf{-.985*} & .418 & -.195 & .184 \\
Condition: AI                                    & \textbf{1.032*} & .448 & .091 & .211 \\
Condition: Combined                              & .107 & .406 & .267 & .194 \\
Self-efficacy $\times$ Data                           & \textbf{-1.549***} & .440 & -.232 & .178 \\
Self-efficacy $\times$ AI                             & \textbf{-1.112**} & .423 & -.105 & .162 \\
Self-efficacy $\times$ Combined                       & -.645 & .399 & -.037 & .165 \\
AI efficacy $\times$ Data                              & \textbf{.871*} & .362 & -.033 & .154 \\
AI efficacy $\times$ AI                                & \textbf{2.202***} & .390 & .198 & .149 \\
AI efficacy $\times$ Combined                          & \textbf{1.173**} & .357 & .008 & .148 \\
\midrule
Log-likelihood                                   & \multicolumn{2}{c}{-1183.0} & \multicolumn{2}{c}{-1312.1} \\
AIC                                              & \multicolumn{2}{c}{2394.0} & \multicolumn{2}{c}{2652.3} \\
Observations                                     & \multicolumn{2}{c}{2880} & \multicolumn{2}{c}{2880} \\
\bottomrule
\end{tabular}
   \begin{tablenotes}
       \item[1] Note: \textit{$^\dagger$p < .10; *p < .05; **p < .01; ***p < .001}
   \end{tablenotes}
\end{center}
\end{table}


The interaction plots in \Cref{fig: interactioneffects} illustrate these dynamics across the different types of discrepancies and conditions, showing how contextual information decreases self-efficacy effects while increasing the effects of AI efficacy.

\begin{figure}[htbp!]
\begin{subfigure}{.32\textwidth}
  \centering
  \includegraphics[width=\textwidth]{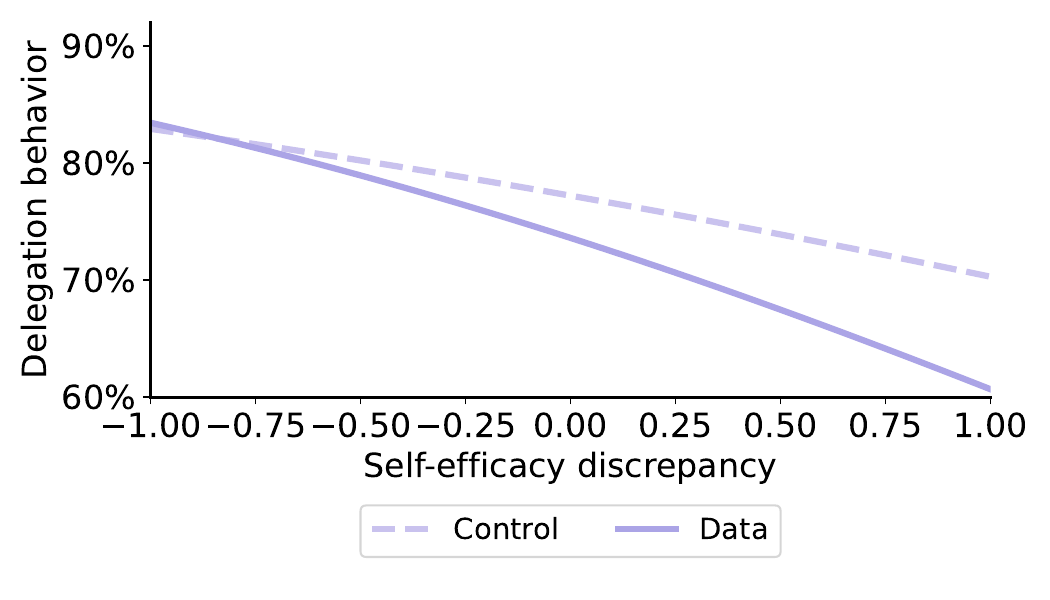}
    \captionsetup{justification=centering}
    \label{subfig: interaction-delegation-se-data}
  \caption{Effect of Data treatment on self-efficacy discrepancy.}
\end{subfigure}
\hfill
\begin{subfigure}{.32\textwidth}
  \centering
  \includegraphics[width=\textwidth]{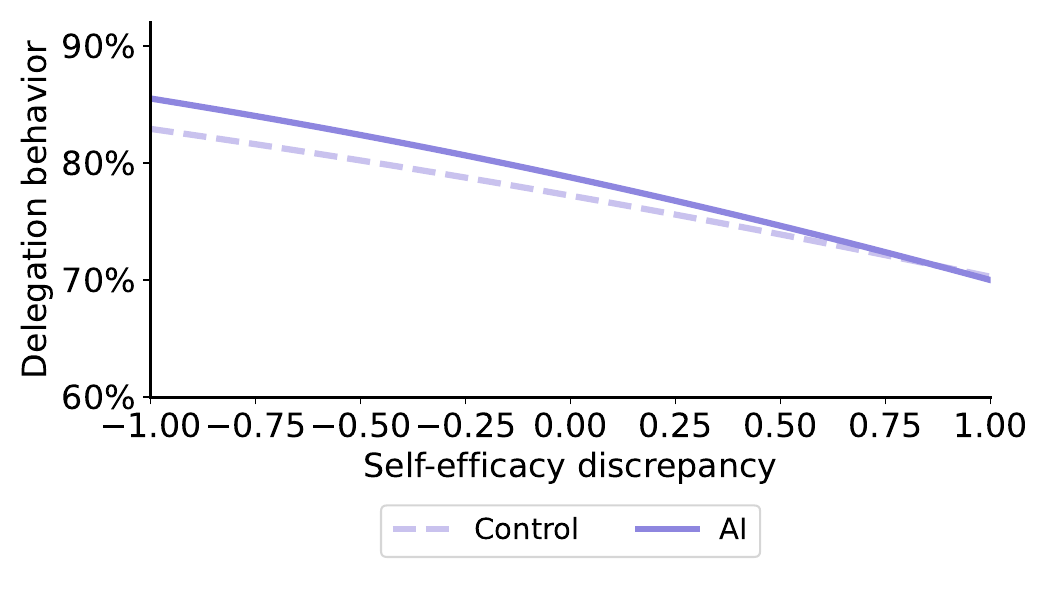}
    \captionsetup{justification=centering}
    \label{subfig: interaction-delegation-se-ai}
  \caption{Effect of AI treatment on self-efficacy discrepancy.}
\end{subfigure}
\hfill
\begin{subfigure}{.32\textwidth}
  \centering
  \includegraphics[width=\textwidth]{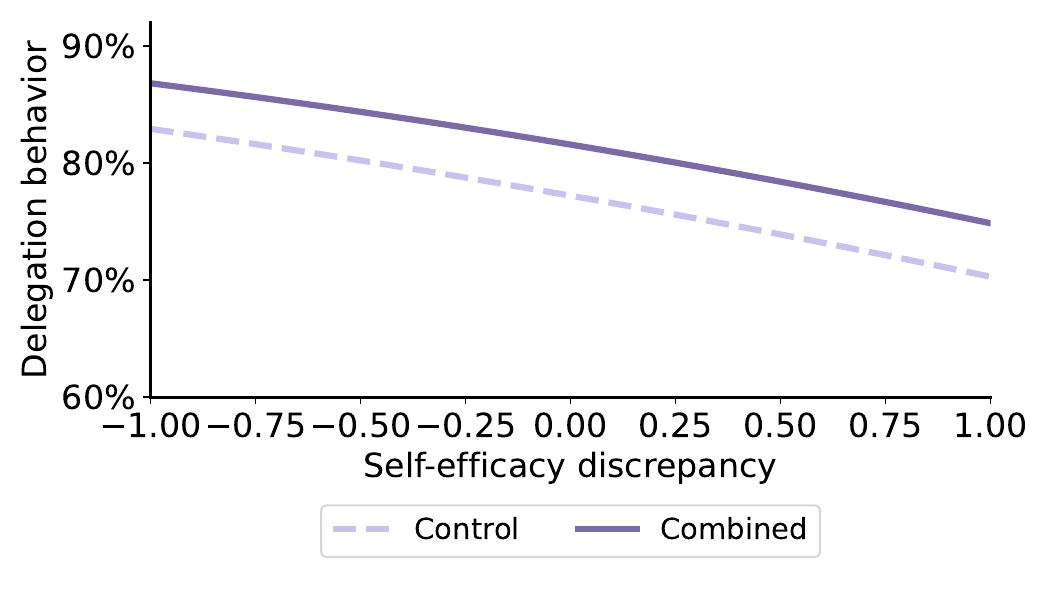}
    \captionsetup{justification=centering}
    \label{subfig: interaction-delegation-se-comb}
  \caption{Effect of combined treatment on self-efficacy discrepancy.}
\end{subfigure}
\captionsetup{justification=raggedright,singlelinecheck=false}
\caption{Interaction effects of the conditions on the relationship of self-efficacy discrepancies and delegation behavior.}
\Description{The plots show the interaction effects of the conditions on the relationship of self-efficacy discrepancies and delegation behavior.}
\label{fig: interactioneffects}
\end{figure}

\begin{figure}[htbp!]
\begin{subfigure}{.32\textwidth}
  \centering
  \includegraphics[width=\textwidth]{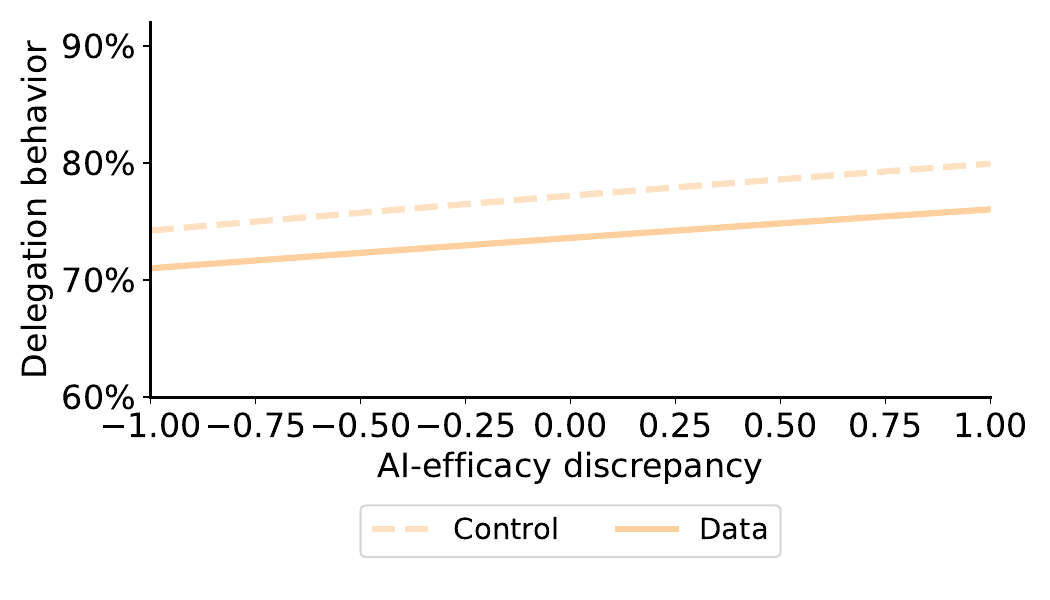}
    \captionsetup{justification=centering}
    \label{subfig: interaction-delegation-ai-data}
  \caption{Effect of data treatment on \\AI efficacy discrepancy.}
\end{subfigure}
\hfill
\begin{subfigure}{.32\textwidth}
  \centering
  \includegraphics[width=\textwidth]{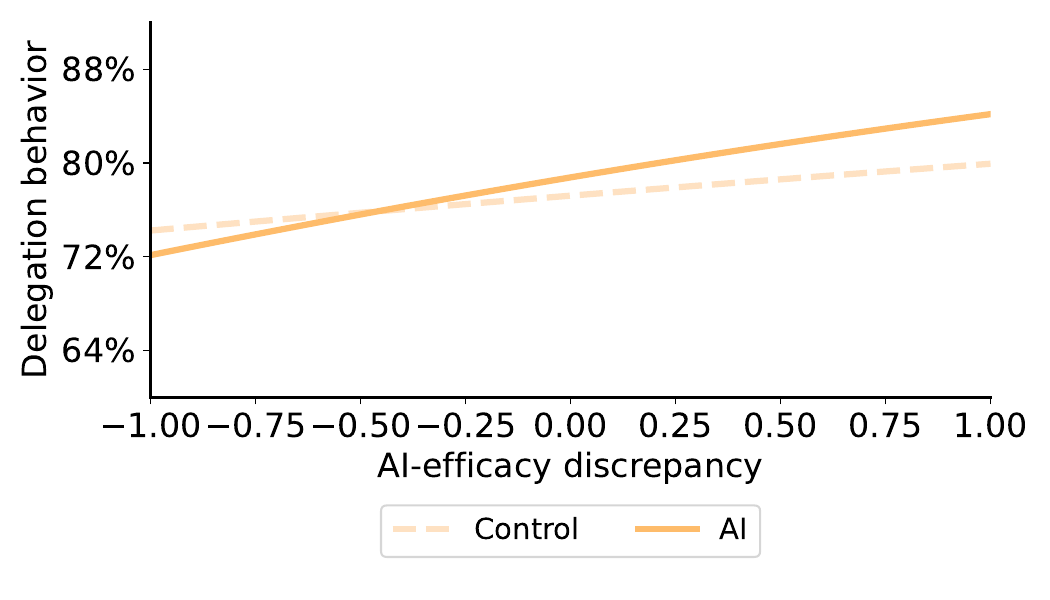}
    \captionsetup{justification=centering}
    \label{subfig: interaction-delegation-ai-ai}
  \caption{Effect of AI treatment on \\AI efficacy discrepancy.}
\end{subfigure}
\hfill
\begin{subfigure}{.32\textwidth}
  \centering
  \includegraphics[width=\textwidth]{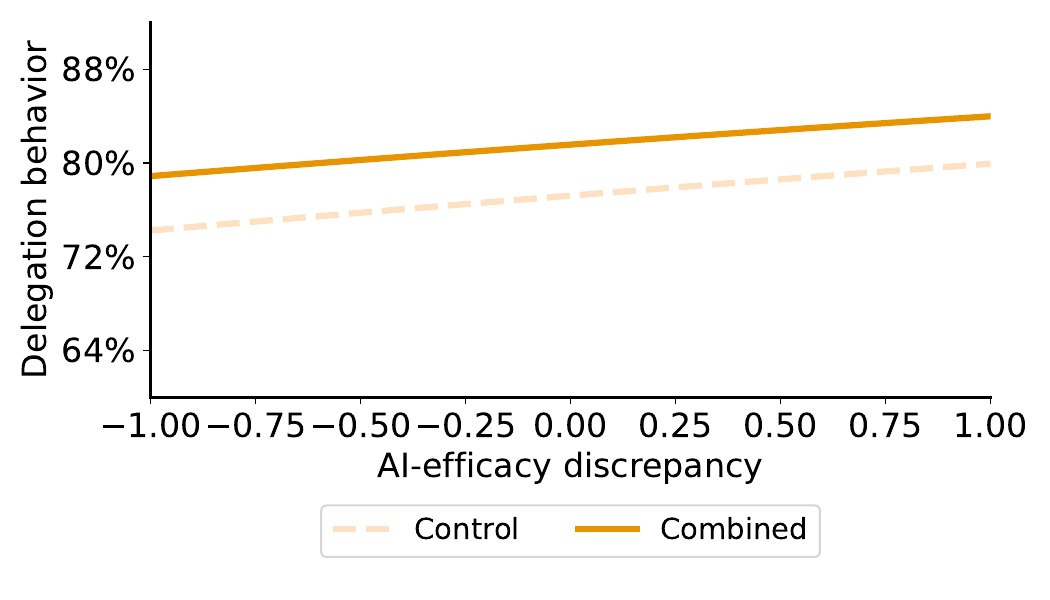}
    \captionsetup{justification=centering}
    \label{subfig: interaction-delegation-ai-comb}
  \caption{Effect of combined treatment \\on AI efficacy discrepancy.}
\end{subfigure}
\captionsetup{justification=raggedright,singlelinecheck=false}
\caption{Interaction effects of the conditions on the relationship of AI efficacy discrepancies and delegation behavior.}
\Description{The plots show the interaction effects of the conditions on the relationship of AI efficacy discrepancies and delegation behavior.}
\label{fig: interactioneffects}
\end{figure}

The performance analysis reveals different dynamics. Self-efficacy discrepancies predict worse team performance ($coef = -.359$, $p < .001$), suggesting that higher instance-wise self-efficacy judgments on specific instances relative to general beliefs lead to worse delegation behavior. AI efficacy discrepancies show only a non-significant trend of a positive effect on performance ($coef = .162$, $p = .073$). Notably, none of the interaction effects reach statistical significance for performance, contrasting with the strong interaction effects observed for delegation behavior.

\begin{findingbox}
    \textbf{Finding 6:} Disconnect between delegation behavior and performance: efficacy discrepancies have larger effects on delegation behavior than on performance outcomes, showing a miscalibration of what drives delegation decisions versus what actually improves the human-AI team performance. Contextual information only affects the impact of efficacy discrepancies on delegation behavior.
\end{findingbox}

\section{Discussion}
\label{sec-discussion}

\subsection{Summary of Results}

This study examines how the relationship between general efficacy beliefs and instance-wise judgments influences delegation behavior and human-AI team performance under varying contextual information conditions. 

First, general efficacy beliefs serve as cognitive anchors for instance-wise judgments, but with asymmetric properties. Self-efficacy judgments aligned closely with general beliefs across all conditions. In contrast, AI efficacy judgments systematically exceeded general beliefs in three of four conditions, creating instance-wise AI optimism. Only AI performance information eliminated this bias, while data and combined information reduced but did not eliminate the overestimation.

Second, contextual information operates asymmetrically as a selective calibrator and broad amplifier. AI performance information specifically eliminated the instance-wise AI optimism bias, while other information types reduced but did not eliminate this systematic overestimation. Simultaneously, contextual information amplified how efficacy discrepancies influenced delegation behavior: data and AI information strengthened the relationship between self-efficacy discrepancies and control retention, while all information types strengthened the relationship between AI efficacy discrepancies and delegation willingness.

Third, efficacy discrepancies showed substantially larger effects on delegation behavior than on performance outcomes. Self-efficacy and AI efficacy discrepancies showed effects of larger magnitude for delegation compared to performance, revealing that humans' delegation strategies systematically diverge from optimal collaboration patterns.


These findings challenge assumptions about transparency in AI systems, suggesting that providing more information may increase humans' efficacy in their delegation behavior without necessarily improving collaborative outcomes. In the following section, we further discuss these findings, relate them to the current discourse in HCI, and derive design guidelines for researchers and practitioners.

\subsection{Implications}
Research on AI-assisted decision-making has extensively examined situational factors such as task difficulty indicators, AI explanation interfaces, and uncertainty visualization to improve collaboration outcomes \citep{zhang2020effect, helldin2013presenting, mcguirl2021supporting, wang2019designing, Holstein2025balancing}.
However, relatively little attention has been paid to how general efficacy beliefs translate into instance-wise judgments and shape behavior. This includes humans' assessments of both their own capabilities and AI competence, and how different contextual information conditions may influence this process. Understanding this dynamic is essential because delegation behavior represents a complex interplay between human expectations, system capabilities, and contextual cues \citep{fugener2019cognitive, pinski2023ai}. Misaligned general beliefs may systematically bias subsequent delegation behavior regardless of the type of contextual information provided, potentially limiting the effectiveness of even well-designed transparency features \citep{bansal2019beyond}, and individual variation in belief-information processing requires more nuanced approaches than one-size-fits-all solutions \citep{miller2019explanation, arrieta2020explainable}.

While our findings relate to the broader literature on trust calibration and confidence in AI systems, our theoretical framing through general efficacy beliefs provides complementary insights that have not been systematically examined in prior work. Trust calibration research examines how reliance aligns with system reliability over time \citep{de2020towards, parasuraman2010complacency}, while confidence calibration focuses on metacognitive accuracy for specific decisions \citep{koriat2012self, ma2024you, von2025knowing}. Our contribution lies in explicitly measuring the temporal transition from general efficacy beliefs to instance-specific efficacy judgments, revealing how these general orientations anchor situation-specific assessments even when transparency mechanisms are present. Prior research measures trust and confidence either as stable pre-task beliefs or as dynamic interaction responses, but rarely examines the relationship between these levels. Our finding that contextual information asymmetrically moderates the translation from general beliefs to instance-wise judgments addresses this gap.

In this work, we take initial steps toward generating insights into the alignment between general efficacy beliefs and instance-wise efficacy judgments, and their relationship with collaborative outcomes. Through our framework, we provide a means to bridge our understanding of efficacy beliefs with actionable insights for improving AI-assisted decision-making. Based on our findings, we derive several implications for research and practice and propose several design guidelines:


\textbf{General efficacy beliefs act as cognitive anchors, but show asymmetric responsiveness to contextual information.} 
Our results demonstrate that both general self-efficacy and AI efficacy beliefs anchor participants' instance-wise judgments, consistent with established research on anchoring bias in human decision-making \citep{tversky1974judgment, bandura1977self}. The moderation patterns reveal an asymmetry. General self-efficacy beliefs consistently predict instance-wise judgments regardless of contextual information, whereas combined contextual information weakens the predictive relationship for general AI efficacy beliefs. This differential responsiveness aligns with research on self-enhancement biases and motivated reasoning, where individuals resist information that challenges their self-understanding while remaining open to updating assessments of external agents \citep{dunning2011dunning, kruger1999unskilled}. 
The psychological mechanisms underlying this asymmetry operate through distinct processes. General self-efficacy beliefs are deeply intertwined with identity and self-concept, activating ego-protective mechanisms that resist downward revisions \citep{vancouver2002two, efferson2020evolution} as it can threaten self-esteem and trigger cognitive dissonance, leading individuals to discount contradictory contextual information \citep{efferson2020evolution}. In contrast, AI efficacy assessments involve evaluating an external tool without identity implications, allowing more flexible updating based on available performance data. Additionally, humans' experience forms stable self-efficacy beliefs, whereas general AI efficacy beliefs are relatively novel and based on limited interaction histories, making them more malleable to new evidence.

The persistence of self-efficacy anchoring suggests that self-protective cognitive processes may limit the effectiveness of contextual information in calibrating self-assessments, even when such information successfully moderates evaluations of AI capabilities. This finding has implications for designing AI-assisted decision-making, as it suggests that interventions targeting self-efficacy may require different approaches than those targeting AI efficacy perceptions \citep{bandura2006guide}. Thus, we propose:

\GuidelineBox{Make Anchoring Bias Visible}{
Help humans recognize their tendency to anchor on initial self-assessments by providing explicit feedback about how their efficacy patterns compare to actual collaboration outcomes over time.
}


\textbf{Asymmetric Anchoring Effects Challenge Instance-Level Design Approaches.} The anchoring effect of general efficacy beliefs has profound implications that fundamentally challenge the prevailing focus of research in AI-assisted decision-making on transparency interventions. Current HCI research predominantly focuses on supporting individual instances \citep{miller2019explanation, helldin2013presenting, Holstein2025balancing}, yet may be addressing symptoms rather than root causes of regularly observed over- and underreliance on AI advice \citep{schemmer2022influence, vaccaro2024combinations}. While these instance-level interventions can inform immediate decisions, they operate within constraints of established general beliefs that bias information interpretation \citep{oeberst2023toward}. Our results demonstrate that even when AI performance information successfully calibrates perceptions of AI efficacy, resistant general self-efficacy beliefs create misaligned delegation strategies, which in turn translate into detrimental human-AI team performance. This finding calls for a paradigmatic shift from designing transparency features for individual decisions to developing interventions that recalibrate underlying general efficacy beliefs. Thus, we propose:

\GuidelineBox{Target Foundational Beliefs, Not Just Decisions}{
Design interventions that address humans' general self-efficacy and AI efficacy beliefs before task engagement to complement transparency approaches for individual AI advice.}


\textbf{Instance-Level AI Optimism and Its Consequences.} 
The finding that participants rated AI capabilities higher on specific instances than their general AI efficacy belief in three out of four conditions reveals a systematic bias we term instance-level ``AI optimism". In delegation contexts, this becomes particularly consequential---unlike decision support settings where humans can partially rely on AI suggestions, delegation requires full commitment based on these potentially inflated instance-wise judgments.

This optimism bias likely stems from multiple cognitive mechanisms \citep{sharot2011optimism}. First, the concreteness of specific instances may trigger availability heuristics, where the salience of immediate task features makes AI capabilities seem more applicable than abstract general assessments suggest \citep{tversky1974judgment}. Second, construal level theory predicts that concrete instances invoke different psychological distance processing than abstract evaluations—specific cases feel more "solvable" and thus AI assistance appears more capable in context \citep{trope2010construal}. Third, the act of considering delegation itself may prime positive expectations about AI performance, as the decision frame shifts from evaluating general AI competence to justifying a specific delegation choice. This combination creates systematic upward bias in instance-wise AI efficacy judgments, even when participants possess accurate general knowledge about AI limitations. This finding challenges transparency-focused approaches that assume providing performance information will automatically calibrate trust \citep{bauer2024mirror}. Instead, we propose:

\GuidelineBox{Expose AI Optimism Patterns}{
Highlight when human's instance-specific AI efficacy systematically exceeds the AI's actual performance to increase awareness of systematic optimism bias.
}


\textbf{General efficacy belief discrepancies expose fundamental challenges in human metacognition for AI-assisted decision-making.} Our results reveal that participants' intuitions about when to delegate to the AI systematically diverge from optimal performance strategies. The difference in effects indicates that the cognitive processes driving delegation behavior are poorly calibrated to actual performance outcomes. The finding that self-efficacy discrepancies harm performance while reducing delegation suggests systematic overconfidence where participants retain control precisely when they should not. Conversely, the weak performance effects of AI efficacy discrepancies despite strong delegation effects suggest that favorable AI assessments do not reliably identify instances where AI delegation would improve outcomes.

This disconnect is evident when comparing our statistical modeling of delegation behavior and performance outcomes: while contextual information created significant interaction effects on delegation behavior, the corresponding interactions for performance outcomes did not reach statistical significance across any condition. While we cannot rule out that smaller performance effects exist, this systematic difference in statistical patterns suggests that the mechanisms driving delegation decisions are more responsive to efficacy discrepancies than the mechanisms determining performance outcomes. This relative dissociation indicates that contextual information changes \textit{how much} people delegate without proportionally affecting \textit{how effectively} they delegate.

This misalignment extends beyond individual biases to system design implications: if humans' natural metacognitive assessments are fundamentally miscalibrated, then AI-assisted decision-making requires mechanisms that actively guide humans toward optimal strategies rather than simply supporting preferences \citep{moore2008trouble, He2023knowing}. The differential effects of contextual information types suggest that information provision may actually exacerbate metacognitive biases by increasing efficacy in already-flawed instance-wise judgment processes. Thus, we propose:

\GuidelineBox{Align Decisions with Outcomes}{
Implement feedback mechanisms that show humans when their reliance behavior (based on efficacy) deviates from optimal performance outcomes, helping calibrate metacognitive assessments over time.
}


\textbf{Contextual Information Creates a Selective Calibration but Broad Amplification Effect.} Our findings reveal an asymmetry in how contextual information operates in AI-assisted decision-making. While contextual AI information eliminated the instance-wise AI optimism bias, all types of contextual information amplified the behavioral impact of efficacy discrepancies on delegation decisions. This creates a situation where calibration is narrow and selective, occurring only for AI efficacy and only with AI performance information, whereas amplification is broad, affecting how all discrepancies translate into delegation behavior across all information conditions. This asymmetry challenges the assumption that providing contextual information uniformly improves decision-making \citep{yu2012influence, efferson2020evolution}. Contextual information may satisfy humans' need for system understanding while making their delegation behavior more responsive to efficacy fluctuations without proportionally improving outcomes.

The amplification mechanism likely operates through increased metacognitive engagement rather than improved accuracy of instance-wise judgments. Contextual information provides reference points that make efficacy discrepancies more salient and actionable. When participants see data distributions or AI performance patterns, deviations from their general beliefs become psychologically more decision-relevant \citep{goddard2004collective}. This heightened salience increases the weight participants place on efficacy fluctuations when making delegation choices, even if the underlying efficacy assessments remain imperfectly calibrated. This explains why behavioral effects increase while performance effects remain weak.

This has implications for AI transparency design, as providing contextual information may inadvertently make delegation behavior more volatile and efficacy-dependent rather than more objectively grounded. Thus, we propose:

\GuidelineBox{Separate Calibration from Decision Support}{
Distinguish between information for understanding (calibration) and information for action (decision support). Provide rich contextual information for system comprehension while offering separate, simpler decision aids for actual delegation choices.
}

Overall, our work extends prior HCI research on AI-assisted decision-making. While previous studies have examined how contextual information on the underlying data \citep{Li2025}, explanations of the AI's reasoning \citep{schoeffer2024explanations}, or performance metrics of the AI across different scenarios \citep{Lu2021, He2023knowing} influence reliance decisions, our contribution lies in systematically distinguishing between general efficacy beliefs formed prior to interaction and instance-wise efficacy judgments made during task execution. Prior work has measured efficacy either statically before tasks \citep{pinski2023ai, Westphal2025} or dynamically during collaboration \citep{chong2023data, li2025confidence}, but has not examined how general beliefs anchor subsequent instance-wise judgments across varying contextual information conditions. Our finding that general efficacy beliefs serve as persistent cognitive anchors—with asymmetric resistance to updating—complements research on cognitive biases in AI-assisted decision-making \citep{rastogi2022deciding}. Similarly, our identification of instance-level AI optimism as a systematic bias distinct from general trust provides a more nuanced understanding of why humans misjudge AI capabilities \citep{romeo2025exploring, vasconcelos2023explanations}. The disconnect between delegation behavior and performance outcomes challenges prevailing emphasis on transparency-focused interventions \citep{miller2019explanation, wang2019designing}. These contributions remain exploratory and context-specific, requiring further validation across diverse settings.

Yet, our findings emerged from collaboration with a deterministic machine learning model with stable performance boundaries. However, the translation from general efficacy beliefs to instance-wise judgments may operate differently in generative AI systems, which exhibit diverse behaviors, produce variable outputs for similar inputs, and lack clearly defined performance profiles. This distinction directly affects our design guidelines: making the anchoring bias visible or exposing AI optimism patterns requires different approaches when system behavior itself is unpredictable---designers cannot show stable performance profiles but must instead convey behavioral tendencies and output variability. Targeting general beliefs may prove less effective when genuine uncertainty about system capabilities exists rather than mere miscalibration. For generative AI systems, designers should therefore prioritize stronger feedback loops in which humans observe delegation outcomes more immediately, support partial reliance rather than binary delegation, and provide instance-level uncertainty indicators.
\section{Limitations and Future Work}
\label{sec-limitations}

Despite the valuable insights gained from this study examining the role of efficacy in human-AI delegation, several limitations must be acknowledged, offering important avenues for future research.

First, our study focuses on a single task domain (income classification) with a specific AI system (decision tree classifier). The decision tree represents a classical machine learning approach with deterministic, interpretable decision boundaries and consistent performance patterns. While this task resembles various real-world scenarios where humans decide whether to rely on AI assistance, the generalizability across different domains, task complexities, and AI capabilities remains uncertain. The income classification task represents a relatively structured prediction problem that may not capture the complexity of real-world collaboration scenarios, such as creative tasks or medical diagnoses. Particularly, generative AI systems exhibit fundamentally different characteristics: non-deterministic outputs, interactivity, unclear performance boundaries, and high variability across similar inputs. These properties may alter how general efficacy beliefs form, how contextual information influences belief-judgment translation, and whether our observed patterns (anchoring effects, AI optimism, selective calibration, and broad amplification) hold. Comparative studies examining how efficacy dynamics vary across classical predictive systems versus generative AI systems would provide crucial insights into the boundary conditions of our findings and inform system-specific design approaches.


Second, while we systematically examined four contextual information conditions, our specific design choices may limit broader applicability. Our contextual information was static, presented once during knowledge enablement, and designed to avoid cognitive overload---yet real-world AI systems often provide dynamic, multi-modal, or continuously updating information. The effectiveness of different information presentations, timing of provision, and adaptive systems that adjust to human behavior represents interesting ground for extending our work. Cross-cultural investigations could explore whether contextual information effectiveness varies across populations with different technological backgrounds, while interactive user interface studies could examine how user-controlled exploration of AI performance information influences general efficacy belief formation and calibration.

Third, our short-term study with twelve delegation instances may not capture how general efficacy beliefs evolve through continuous assistance by AI, particularly as no feedback on the collaborative performance was available. This brief engagement limits our understanding of whether general belief anchoring effects diminish with experience, how learning from delegation outcomes influences subsequent assessments, or how humans develop sophisticated mental models over time \citep{spitzer2024don, bansal2019beyond}. Longitudinal designs tracking general efficacy beliefs and delegation patterns would illuminate whether our observed cognitive biases represent stable individual differences or unstable responses that calibrate through experience. Such extended studies could also examine how general efficacy beliefs transfer across different AI systems and whether humans develop generalizable collaboration strategies that improve performance across diverse human-AI partnerships.


Finally, our focus on task accuracy as the primary performance measure may not fully capture the multifaceted nature of factors that determine effective collaboration \citep{sivaraman2025over}. Real-world AI-assisted decision-making often involves trade-offs between efficiency, human satisfaction \citep{Westphal2025}, learning outcomes, and long-term skill development that binary delegation paradigms cannot address. Developing comprehensive frameworks that account for multiple performance dimensions and collaboration patterns beyond delegation would enrich our understanding of when and how general efficacy beliefs optimize human-AI teamwork. Research incorporating measures of cognitive load, human confidence development, and skill transfer could reveal whether delegation behavior that appears suboptimal for immediate accuracy serves valuable insights into AI-assisted decision-making, informing the design of systems that balance short-term performance with sustainable partnership effectiveness.
\section{Conclusion}
\label{sec-conclusion}

This study examines how general efficacy beliefs translate into instance-wise judgments in AI-assisted decision-making, investigating the moderating role of contextual information and consequences for delegation behavior and human-AI team performance. Our findings reveal the nuanced interplay between self-efficacy, AI efficacy, and contextual information, which shapes delegation behavior in ways that diverge from enhanced performance outcomes. Contextual information operates asymmetrically: while it calibrates instance-wise efficacy judgments (eliminating AI optimism bias) and reduces anchoring effects, it simultaneously amplifies how efficacy discrepancies drive delegation behavior without proportionally improving collaborative outcomes. This disconnect between humans' intuitive delegation strategies and actual collaboration effectiveness underscores the need for AI design that goes beyond transparency to actively guide users toward calibrated reliance decisions. Understanding how general efficacy beliefs and instance-wise judgments, together with contextual information, shape human-AI delegation provides a foundation for developing collaborative systems that leverage human efficacy while correcting for systematic biases that hinder effective teamwork. Ultimately, we propose design guidelines for the development of collaborative systems to enable partnerships that increase the use of both human agency and AI capabilities.


\section*{Acknowledgements}

Generative AI tools were utilized throughout this work. Specifically, ChatGPT, Claude, and GitHub Copilot were employed to generate code for visualizations. Additionally, ChatGPT, DeepL Write, and Grammarly were used to enhance the writing quality of tutorials and explanations provided to participants during the experiments, as well as to improve the language across all sections of this paper.
\bibliographystyle{ACM-Reference-Format}
\bibliography{references}

\newpage
\appendix
\section{Appendix}
\label{sec-appendix}

\begin{figure}[htbp!]
  \centering
  \begin{minipage}{\linewidth}
    \centering
    \begin{subfigure}[t]{.48\linewidth}
      \centering
      \includegraphics[width=\linewidth]{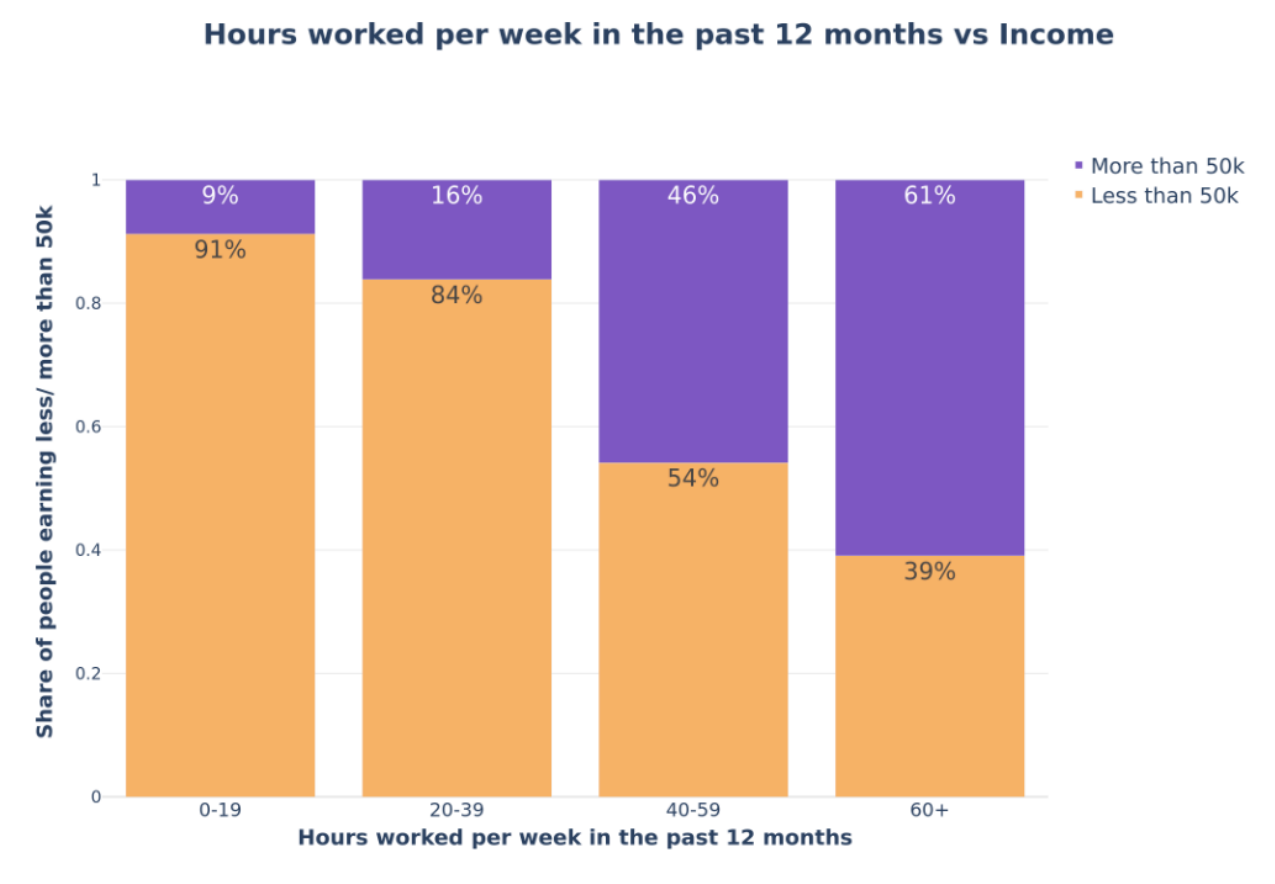}
      \caption{Data information.}
    \end{subfigure}
    \hfill
    \begin{subfigure}[t]{.40\linewidth}
      \centering
      \includegraphics[width=\linewidth]{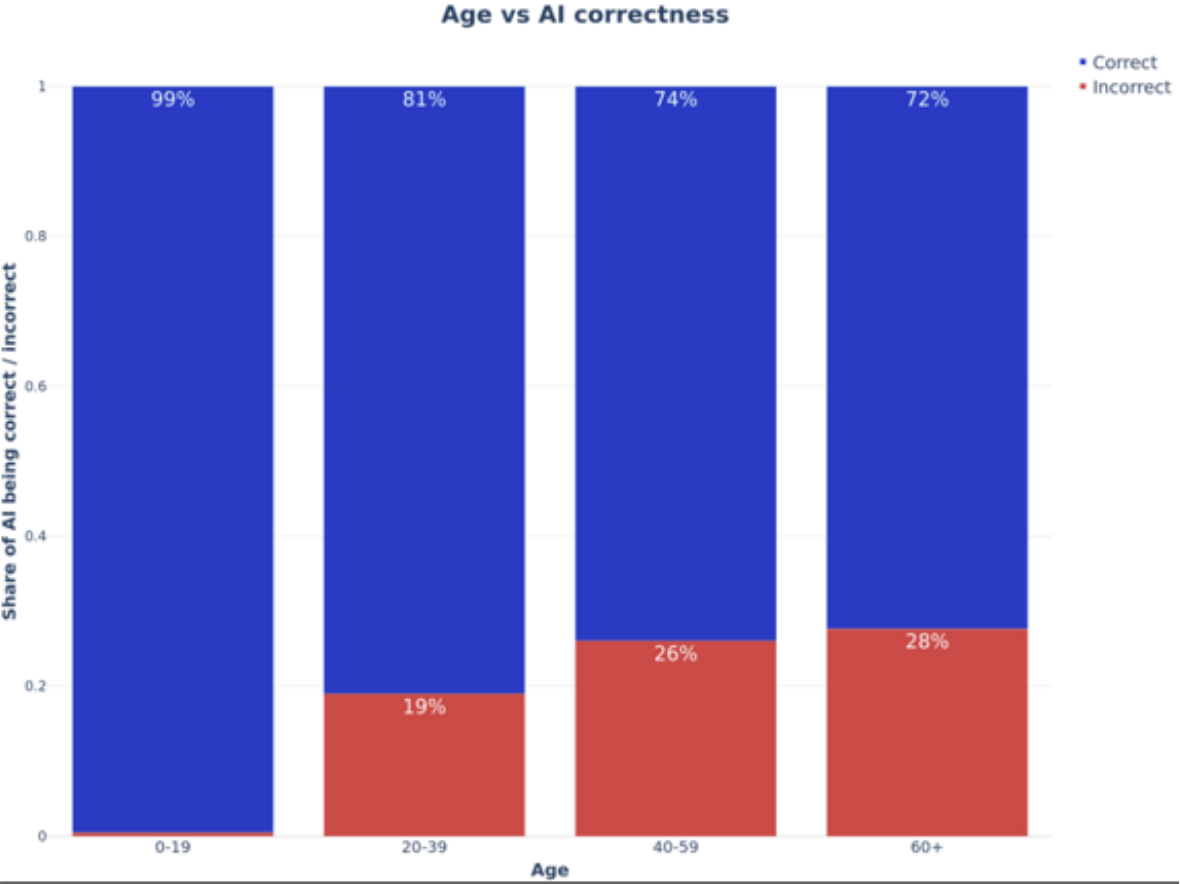}
      \caption{AI information.}
    \end{subfigure}
  \end{minipage}
  \caption{Examples of contextual information.}
  \label{fig: contextualinformation}
\end{figure}

\begin{table}[htbp]
\centering
\caption{Overview of All Questionnaire Items and Measures}
\Description{The table shows an Overview of All Questionnaire Items and Measures.}
\label{tab:all_measures}
\small 
\begin{tabular}{p{2cm}p{1.2cm}p{8cm}p{1.5cm}}
\toprule
\textbf{Construct} & \textbf{Response Format} & \textbf{Item Text} & \textbf{Reference}\\
\midrule
\multicolumn{4}{l}{\textbf{Pre-Task Measures}} \\
\midrule
General Self-Efficacy & 7-point Likert & \makecell[l]{1. I am confident about my ability to do the task. \\ 2. I have mastered the skills necessary for the task. \\ 3. I am self-assured about my capabilities to perform the task.} & \citet{spreitzer1995psychological}  \\
\addlinespace
General AI Efficacy & 7-point Likert & \makecell[l]{1. I am confident in the AI's ability to do the task. \\ 2. The AI has the required skills for the task. \\ 3. I am convinced of the AI's capabilities to perform the task.} & \citet{spreitzer1995psychological} \\
\midrule
\multicolumn{4}{l}{\textbf{Instance-Level Measures}} \\
\midrule
Instance Self-Efficacy & 0-100\% Slider & How well are you suited to solve this task? & \citet{pinski2023ai}\\
\addlinespace
Instance AI Efficacy & 0-100\% Slider & How well is the AI suited to solve this task? & \citet{pinski2023ai}\\
\addlinespace
Perceived Difficulty & 6-point Likert & How difficult do you rate this instance? & Self-developed\\
\midrule
\multicolumn{4}{l}{\textbf{Post-Task Measures}} \\
\midrule
AI Literacy & 5-point Likert & When it comes to Artificial Intelligence (AI), I believe I have... & \citet{ehsan2021explainable} \\
\addlinespace
Task Familiarity & 6-point Likert & How familiar are you with the task domain of classifying people's income? & Self-developed\\
\bottomrule
\end{tabular}
\end{table}

\begin{table}[htbp]
\centering
\caption{Shapiro–Wilk normality test \textit{p}-values for the four conditions}
\label{tab:normality-pvals}
\setlength{\tabcolsep}{12pt}  
\begin{tabular}{lrr}
\toprule
\textbf{Condition} & \makecell{\textbf{Self‑Efficacy} \\ \textbf{Discrepancies}} & \makecell{\textbf{AI‑Efficacy} \\ \textbf{Discrepancies}} \\
\midrule
Control  & .0797 & .3800 \\
Data     & .8491 & .7525 \\
AI       & .5016 & .0138 \\
Combined & .0286 & .0270 \\
\bottomrule
\end{tabular}
\end{table}

\begin{table}[htbp]
  \centering
  \caption{Self‑efficacy: discrepancy between general beliefs and instance‑wise judgments.}
  \label{tab:self-efficacy-discrepancy}
  \begin{tabular}{lrrrrr}
    \toprule
    Condition & $\Delta$ & $t(59)$ & $p$ & $q$ & $d$ \\
    \midrule
    Control   & .027  & 1.24  & .220 & .654 & .16 \\
    Data      & -.012 & -.72 & .475 & .654 & -.09\\
    AI        & -.015 & -.69 & .491 & .654 & -.09\\
    Combined  & -.010 & -.41 & .682 & .682 & -.05\\
    \bottomrule
  \end{tabular}
  \begin{tablenotes}
    \item[] Note: No significant discrepancies from zero were observed.
  \end{tablenotes}
\end{table}

\begin{table}[htbp]
  \centering
  \caption{AI‑efficacy: discrepancy between general AI efficacy beliefs and instance‑wise judgments.}
  \label{tab:ai-efficacy-discrepancy}
  \begin{tabular}{lrrrrr}
    \toprule
    Condition & $\Delta$ & $t(59)$ & $p$ & $q$ & $d$ \\
    \midrule
    Control   & .098 & 4.60 & $<$.001*** & $<$.001 & .59 \\
    Data      & .065 & 3.48 & .001** & .001 & .45 \\
    Combined  & .085 & 3.77 & $<$.001*** & $<$.001 & .49 \\
    AI        & .030 & 1.46 & .151 & .151 & .19 \\
    \bottomrule
  \end{tabular}
  \begin{tablenotes}
    \item[] Note: $^\dagger p < .10$; $*p < .05$; $**p < .01$; $***p < .001$
  \end{tablenotes}
\end{table}

\begin{figure}[htbp!]
  \centering
  \includegraphics[width=.45\textwidth]{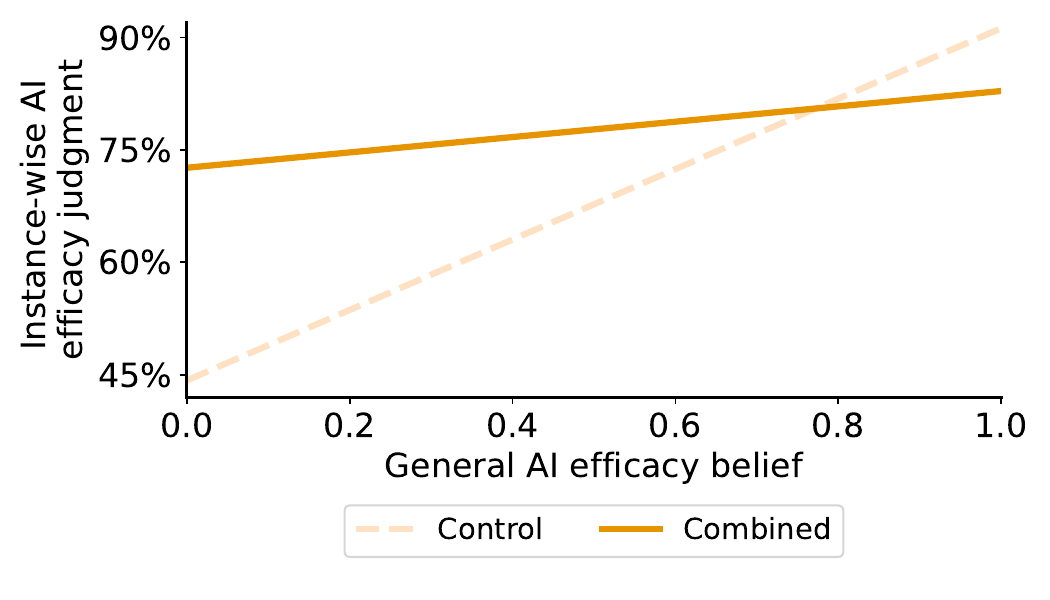}
    \captionsetup{justification=centering}
    \centering
\captionsetup{justification=raggedright,singlelinecheck=false}
\caption{Interaction effect of contextual combined information on general AI efficacy belief on instance-wise AI efficacy judgment.}
\Description{The plot shows the interaction effect of contextual combined information on AI efficacy belief on AI efficacy judgment as two line plots.}
\label{fig: aiefficacy-interaction}
\end{figure}

\end{document}